\documentclass [12pt]{article}
\usepackage{amsfonts,amsmath,amssymb}
\usepackage{graphicx}
\usepackage{epic,eepic,color,graphicx}

\newfont{\twelvemsb}{msbm10 scaled\magstep1}
\newfont{\eightmsb}{msbm8}
\newfam\msbfam
\textfont\msbfam=\twelvemsb \scriptfont\msbfam=\eightmsb
\catcode`\@=11
\def\Bbb{\ifmmode\let\next\Bbb@\else
\def\next{\errmessage{Use \string\Bbb\space only in math mode}}\fi\next}
\def\Bbb@#1{{\fam\msbfam{{#1}}}}

\newcommand{\be}{\begin{equation}}
\newcommand{\ee}{\end{equation}}
\newcommand{\ba}{\begin{eqnarray}}
\newcommand{\ea}{\end{eqnarray}}

\begin{document}

\sloppy
\renewcommand{\thefootnote}{\fnsymbol{footnote}}
\newpage
\setcounter{page}{1} \vspace{0.7cm}
\begin{flushright}
05/08/15
\end{flushright}
\vspace*{1cm}
\begin{center}
{\bf Reciprocity and self-tuning relations without wrapping}\\
\vspace{1.8cm} {\large Davide Fioravanti $^a$, Gabriele Infusino $^b$ and
Marco Rossi $^c$
\footnote{E-mail: fioravanti@bo.infn.it, gabriele.infusino@fis.unical.it, rossi@cs.infn.it}}\\
\vspace{.5cm} $^a$ {\em Sezione INFN di Bologna, Dipartimento di Fisica e Astronomia,
Universit\`a di Bologna, \\
Via Irnerio 46, Bologna, Italy} \\
\vspace{.3cm} $^b${\em Dipartimento di Fisica dell'Universit\`a
della Calabria, I-87036
Arcavacata di Rende, Cosenza, Italy and Laboratoire Jean Alexandre Dieudonn\'e, Universit\'e Nice Sophia Antipolis, 06100 Nice, France} \\
\vspace{.3cm} $^c${\em Dipartimento di Fisica dell'Universit\`a
della Calabria and INFN, Gruppo collegato di Cosenza, I-87036
Arcavacata di Rende, Cosenza, Italy} \\
\end{center}
\renewcommand{\thefootnote}{\arabic{footnote}}
\setcounter{footnote}{0}
\begin{abstract}
{\noindent We consider scalar Wilson operators of ${\cal N}=4$ SYM at high spin, $s$, and generic twist in the multi-color limit. We show that the corresponding (non)linear integral equations (originating from the asymptotic Bethe Ansatz equations) respect certain 'reciprocity' and functional 'self-tuning' relations up to all terms $\frac{1}{s(\ln s)^n}$ (inclusive) at any fixed 't Hooft coupling $\lambda$. Of course, this relation entails straightforwardly the well-known (homonymous) relations for the anomalous dimension at the same order in $s$. On this basis we give some evidence that wrapping corrections should enter the non-linear integral equation and anomalous dimension expansions at the next order $\frac{(\ln s)^{2}}{s^2}$, at fixed 't Hooft coupling, in such a way to re-establish the aforementioned relation (which fails otherwise).}
\end{abstract}
\vspace{4cm}

\newpage

\section{Introduction, aims and results}
\setcounter{equation}{0}

One of the major achievements of modern theoretical physics is the so-called AdS/CFT correspondence \cite{MGKPW1,MGKPW2,MGKPW3} and its description in terms of integrability tools 
\cite{MZ,BS1,BS2, BS3,BS4,BS5,BES,WRA1,WRA2,TBA1,TBA2,TBA3,TBA4,TBA5,Y,CFT}. In fact, being a strong/weak coupling duality the non-perturbative, exact --  thought not necessarily {\it explicit} \footnote{As a simple example we can mention, just with reference to the present paper, that the following non-linear integral equation (which governs the spectrum) is not {\it explicitly solvable}.} -- nature of integrability is of incomparable value and utility. In particular, the spectrum of anomalous dimensions of composite operators in ${\cal N}=4$ super Yang-Mills (SYM) theory ought to correspond to the energy spectrum of states in type IIB superstring theory in $\text{AdS}_5\times\text{S}^5$, and both must be described by an integrable system.

Among the different sectors of multi-color ${\cal N}=4$ SYM (perturbatively closed under renormalisation), one of the most studied ones is the so-called $sl(2)$ scalar twist sector. This is spanned by local composite operators of single trace form
\begin{equation}
{\mbox {Tr}} ({\cal D}^s {\cal Z}^L)+.... \, , \label {sl2op}
\end{equation}
where ${\cal D}$ is a (light-cone) covariant derivative acting in all the possible ways on the $L$ complex bosonic fields ${\cal Z}$, the trace ensuring gauge invariance. The Lorentz spin of these operators is $s$ and $L$ coincides with the twist, {\it i.e.} the classical dimension minus the spin.
The AdS/CFT correspondence relates operators (\ref {sl2op}) to spinning folded closed strings
on $\text{AdS}_5\times\text{S}^5$ spacetime, with $\text{AdS}_5$ and $\text{S}^5$ angular momenta $s$ and $L$, respectively \cite{GKPII,FT}.

One of the several reasons for the large interest in these operators is their similarity with twist operators in QCD, where, maybe, the scalars are substituted by fermions, {\it i.e.} the quarks, or gauge fields: because of integrability in ${\cal N}=4$ these cases would be dealt with in an analogous manner \cite{FPR1,FPR2}. Similarities among the two theories give the possibility to believe that QCD  could take many advantages of a full all-loop solution of its supersymmetric counterpart. In QCD, in the framework of Partonic Model, the Lorentz spin $s$ is the conjugated variable, in the Mellin transform (of the splitting function, for instance, which gives the anomalous dimension), to the Bjorken variable $x$, namely the fraction of the hadron momentum carried by the single parton \footnote{Of course, the coupling does run in QCD, unlike what happens in the maximally supersymmetric theory}. In this context, two regimes emerge naturally: $x\rightarrow 0$, governed by the BFKL equations \cite{BFKL} and $x\rightarrow 1$, corresponding exactly to large values of the Lorentz spin, $s\rightarrow\infty$. Properties of this second (called \emph{quasi-elastic}) regime can be deduced by large spin results in three loops twist-2 QCD calculations. In particular, we can enlighten two main features about anomalous dimension of twist operators:
\begin{enumerate}
\item The leading term has a logarithmic scale
\begin{equation}
\gamma(s)\sim \ln s,\;\; s\rightarrow\infty \, .
\end{equation}
\item Sub-leading terms obey hidden relations, the Moch-Vermaseren-Vogt constraints \cite {MVV1,MVV2}: in brief, terms proportional to $\ln s /s$ and $1/s$ are completely determined by terms proportional to $\ln s$ and $s^0$. These constraints are related with spacetime reciprocity of deep inelastic scattering and its crossed version of $e^+ e^-$ annihilation into hadrons.
\end{enumerate}
${\cal N}=4$ gauge theory shares at large $s$ these features, and, besides, allows us an understanding of their origin and thus possible extension to QCD. In specific, the asymptotic large $s$ series of the anomalous dimensions are believed to be constrained by non-perturbative (in $\lambda =8\pi^2 g^2$, the 't Hooft multi-color coupling) functional relations that work for any finite value of the twist $L$. To be more precise, conformal symmetry implies that anomalous dimensions $\gamma (g,L,s)$ of twist operators are functions of the conformal spin: this translates into the following 'self-tuning' functional relation \cite {DMS1,DMS2}:
\be
\gamma (g,L,s)= P \left (s+\frac{1}{2}\gamma (g,L,s)\right ) \, . \label {self-tun}
\ee
Additionally, this has to be meant asymptotically in the sense that the function \footnote {The function $P$ in (\ref
{self-tun}) actually depends on the twist of the operator as well.}
\be
P(s)=\sum _{n=0}^{\infty} \frac{a^{(n)}(\ln C(s))}{C(s)^{2n}} \, , \label {reci}
\ee
is represented by a series in $s$ via the conformal Casimir
\be
C(s)^2=\left (s+\frac{L}{2}-1 \right) \left (s+\frac{L}{2}\right ) \,  \label {spinC}
\ee
only. Relation (\ref {reci}) is equivalent to the so-called 'reciprocity symmetry' $x \rightarrow -x$, but for the Mellin space variable $s$, and an important information is that the function $a^{(n)}$ has the form of an upper truncated Laurent series:
\be
a^{(n)}(\ln C(s))=\sum _{m=-\infty}^{M} b^{(n)}_m(g,L) (\ln C(s))^m \label {an-exp} \, , \quad M < \infty \, ,
\ee
{\it i.e.} $a^{(n)}(\ln C(s))$ depends on $s$ only through powers of $\ln C(s)$.
For twist two and three negative powers of $\ln C(s)$ are absent and $a^{(n)}$ is a polynomial;
for generic twist, however, one has to cope with the infinite Laurent series (\ref {an-exp}).

Relations (\ref {self-tun}, \ref {reci}), both in QCD and in ${\cal N}=4$ SYM, are developed, checked in various cases and discussed in \cite{DMS1,DMS2,REC1,REC2,REC3,REC4,REC5,REC6,REC7,REC8,REC9}. Recently, they have been proven, restrictively to twist two operators, but in a generic conformal field theory, in \cite {ABL}, with some arguments for their validity in non-conformal theories at the end. Clearly, they provide important information on the high spin expansion of anomalous dimension of twist operators. Unlike QCD, in ${\cal N}=4$ SYM it is possible to obtain better and more suitable results as we can consider these relations into the framework of integrability. The latter was firstly discovered in the planar limit for the purely bosonic $so(6)$ sector at one loop \cite {MZ}; then it was extended to all the gauge theory sectors and to all loops \cite{BS1,BS2,BS3,BS4,BS5,BES}. In specific, it was found that every composite operator can be thought of as a state of a 'spin chain', whose hamiltonian is the dilatation operator itself, although the latter does not have an explicit expression of the spin chain form, but for the first few loops. Nevertheless, the spectrum of infinitely long operators has turned out to be exactly described by a set of Asymtoptic Bethe Ansatz (ABA) equations \cite{BS1,BS2,BS3,BS4,BS5,BES}. On the other hand, anomalous dimensions of operators with finite quantum numbers depend not only on ABA data but also on finite size 'wrapping' corrections \cite {WRA1,WRA2}. Subsequent progress has shown that a set of Thermodynamic Bethe Ansatz (TBA) equations \cite{TBA1,TBA2,TBA3,TBA4,TBA5} or an equivalent $Y$-system of functional equations \cite{Y}, together with certain additional information \cite{CFT}, provides a solid ground for exact (any length, any coupling) predictions on anomalous dimensions of planar ${\cal N}=4$ SYM.

Despite this impressive progress, we believe that it is still important to define the largest domain of composite operators for which the 'simpler' ABA equations give the correct anomalous dimensions, especially in connexion with other well-established relevant equations. In fact, we intended this to be the main aim of this paper and the most natural setting to perform this study to be the reformulation of ABA equations as one (Non)linear Integral Equation (NLIE) \cite{BFR}.

Generically and sketchily, for operators composed of $L$ elementary fields ABA gives the correct perturbative expansion of the anomalous dimension up to $L-1$ loops. Starting from $L$ loops, 'wrapping' diagrams - which are not taken into account by ABA - start to contribute. In this general framework, the high spin limit of fixed twist operators seems to offer
a better scenario. Perturbative (up to six loops) computations \cite {BJL1,BJL2,BJL3} for short (twist two and three) operators show that wrapping diagrams (which enter from four loops on) actually give contributions which in the high spin limit behave as $O\left (\frac{(\ln s)^2}{s^2} \right )$. It is then natural to ask if such property extends to higher (and possibly to all) orders of perturbation theory.
In this paper we want to provide evidence in favour of this picture, by using the
self-tuning and reciprocity properties.
In order to do that, we first rewrite (Section 2) the ABA equations as NLIEs for the counting function. Then, in Section 3 we specialise ourselves to the minimal anomalous dimension state and go to the high spin limit, while keeping the twist finite: upon computing the positions of the external holes and the effect of the non-linear terms, we write a linear integral equation equivalent to ABA up to the orders $\frac{1}{s (\ln s )^n}$, $n\in \mathbb{Z}$, $n\geq -1$ (inclusive). In Section 4, we use this linear integral equation to compute at the same order of $s$, but at all values of the coupling, the ABA prediction for the minimal anomalous dimension. Then, in Section 5 we show the latter to satisfy the self-tuning and reciprocity relations. Interestingly, we also find that the solution of the linear integral equation respects suitable self-tuning and reciprocity relations (up to this order in $s$). Finally, we provide some arguments supporting the idea that at high spin wrapping corrections affect twist operators starting from orders
$\frac{(\ln s )^2}{s^2}$, so that self-tuning and reciprocity relations still hold (and likely also a modified (non)linear integral equation).

\section{From the ABA to the NLIE}
\setcounter{equation}{0}

As planned in the introduction, we start from the ABA equations \cite {BS1,BS2,BS3,BS4,BS5,BES} for the $sl(2)$ sector of  ${\cal N}=4$ SYM,
\begin{equation}
\left ( \frac {u_k+\frac {i}{2}}{u_k-\frac {i}{2}} \right )^L \left
( \frac {1+\frac {g^2}{2{x^-(u_k)}^2}}{1+\frac {g^2}{2{x^+(u_k)}^2}}
\right )^L=\mathop{\prod^s_{j=1}}_{j\neq k}  \frac
{u_k-u_j-i}{u_k-u_j+i}  \left ( \frac {1-\frac
{g^2}{2x^+(u_k)x^-(u_j)}}{1-\frac {g^2}{2x^-(u_k)x^+(u_j)}} \right )^2
e^{2i\theta (u_k,u_j)}\, , \label {manyeq}
\end{equation}
where
\begin{equation}
x^{\pm}(u_k)=x(u_k\pm i/2) \, , \quad x(u)=\frac
{u}{2}\left [ 1+{ \sqrt {1-\frac {2g^2}{u^2}}} \right ] \, , \quad
\lambda =8\pi ^2 g^2 \, ,
\end{equation}
$\lambda $ being the 't Hooft coupling.
The so-called dressing factor \cite {AFS1,AFS2,BES} $\theta (u,v)$ is given by
\begin{equation}
\theta (u,v)=\sum _{r=2}^{\infty}\sum _{\nu =0}^{\infty} \beta
_{r,r+1+2\nu}(g)
[q_r(u)q_{r+1+2\nu}(v)-q_r(v)q_{r+1+2\nu}(u)] \, ,
\end{equation}
the functions $\beta _{r,r+1+2\nu}(g)=g^{2r+2\nu-1}2^{1/2-r-\nu}c_{r,r+1+2\nu}(g)$ being
\begin{eqnarray}
\beta _{r,r+1+2\nu}(g)&=&2 \sum _{\mu =\nu}^{\infty} \frac {g^{2r+2\nu+2\mu}}{2^{r+\mu+\nu}} (-1)^{r+\mu+1}\frac {(r-1)(r+2\nu)}{2\mu +1} \cdot \nonumber \\
&\cdot & \left ( \begin{array}{cc} 2\mu +1 \\ \mu -r-\nu+1
\end{array} \right )\left ( \begin{array}{cc} 2\mu +1 \\ \mu -\nu
\end{array} \right )              \zeta (2\mu +1)
\end{eqnarray}
and $q_r(u)$,
\begin{equation}
q_r(u)=\frac {i}{r-1} \left [ \left (\frac {1}{x^+(u)}\right
)^{r-1}-\left (\frac {1}{x^-(u)}\right )^{r-1} \right ] \, ,
\end{equation}
being the expression of the $r$-th charge in terms of the rapidity $u$.
Operators (\ref {sl2op}) of twist $L$ correspond to zero momentum states of the $sl(2)$ spin chain
described by an even number $s$ of real Bethe roots $u_k$ which satisfy (\ref  {manyeq}).
For a state described by the set of Bethe roots $\{ u_k \}, k=1, \ldots, s$, the eigenvalue of the $r$-th charge is
\be
Q_r(g,L,s)=\sum _{k=1}^s q_r(u_k) \,  \label {Qr} .
\ee
In particular, (asymptotic) anomalous dimension of (\ref {sl2op}) is
\be
\gamma  (g,L,s)=g^2 Q_2 (g,L,s)\, .
\ee
Let us focus (in this Section; from Section 2 on, we will restrict to the minimal anomalous dimension state) on states described by positions of roots which are
symmetric with respect to the origin. These are in particular zero momentum states.
An efficient way to treat states described by solutions to a  (possibly large) number of (algebraic) Bethe Ansatz equations consists in writing one non-linear integral equation completely equivalent to them ({\it cf.} \cite{NLIE} and references therein for the idea without holes degree of freedom). The nonlinear integral equation is satisfied by the counting function $Z(u)$, which in the case (\ref {manyeq})
reads as
\be
Z(u)=\Phi (u)-\sum _{k=1}^{s}
\phi (u,u_k) \, , \label {Z}
\ee
where
\be
\Phi (u)=\Phi _0(u)+\Phi _H (u) \, , \quad \phi (u,v)=\phi _0(u-v)+\phi _H (u,v) \, ,
\ee
with
\ba
\Phi _0(u) &=&-2L \arctan 2u \, , \quad \Phi _H(u)=-iL \ln \left ( \frac {1+\frac {g^2}{2{x^-(u)}^2}}{1+\frac {g^2}{2{x^+(u)}^2}} \right )\, ,  \label {Phi} \\
\phi _0(u-v)&=& 2\arctan (u-v) \, ,\quad \phi _H(u,v)=-2i \left [ \ln \left ( \frac {1-\frac {g^2}{2x^+(u)x^-(v)} }{1-\frac {g^2}{2x^-(u)x^+(v)}} \right )+i\theta (u,v)\right] \, . \label {phi}
\ea
It follows from its definition that the counting function $Z(u)$ is a monotonously decreasing function. In addition, in the limit $u \rightarrow \pm \infty$, since
\be
\phi (u,v)+\phi (u,-v) \rightarrow  \pm 2\pi - \frac{4}{u}+\frac{2ig^2}{u}\left (\frac{1}{x^{-}(v)}-\frac{1}{x^{+}(v)} \right )+
O\left (\frac{1}{u^3}\right ) \, ,
\ee
one has the asymptotic behaviour
\be
u \rightarrow \pm \infty \, , \quad Z(u)\rightarrow  \mp(L+s)\pi + \frac{L+2s+\gamma (g,L,s)}{u}+
O\left (\frac{1}{u^3}\right ) \, .  \label {Zlimit}
\ee
This means that there are $L+s$ real points $\upsilon _k$ such that $e^{iZ(\upsilon _k)}=(-1)^{L+1}$.
It is a simple consequence of the definition of $Z(u)$ that $s$ of them coincide with the Bethe roots $u_k$. For Bethe equations
(\ref {manyeq}) Bethe roots are all real and are all contained in an interval $[-b,b]$ of the real line.
The remaining $L$ points are called 'holes' \cite{BGK,FRS1,FRS2,FRS3,FRS4,FRS5,FRS6,BFR}, they also are real and they will be denoted as $x_h$.
One should distinguish between $L-2$
'internal' or 'small' holes $x_h, h=1,...,L-2$, which reside inside the interval $[-b,b]$, and two 'external' or 'large' holes $x_{L-1}=-x_L$, with $x_L>b$.

We finally remark that anomalous dimension appears (\ref {Zlimit}) in the limit $u \rightarrow \infty$ of the counting function. We will come back on this fact in Appendix \ref {kotlipgen}.

\medskip

As we are in presence of holes, we may follow the extension of the idea as developed in \cite {FMQR} and make use of the Cauchy theorem to obtain a simple integral formula ($Z'(v)=\frac{d}{dv}Z(v)$; {\it cf.} also \cite{FRXYZ} for more details on the following formul\ae)
\be
\sum _{k=1}^{s} O(u_k)+\sum _{h=1}^{L} O(x_h)=-\int _{-\infty}^{+\infty} \frac {dv}{2\pi} O(v)
Z^{\prime}(v)+  \int _{-\infty}^{+\infty}\frac {dv}{\pi}
O(v)\frac {d}{dv} {\mbox {Im}} \ln
[1+(-1)^{L} \, e^{iZ(v-i0^+)}] \, . \label {cauchy}
\ee
Application of (\ref {cauchy}) to the derivative of (\ref {Z}) gives
\ba
Z'(u)&=&\Phi '(u) + \int _{-\infty}^{+\infty} \frac {dv}{2\pi} \frac{d}{du}\phi (u,v)
Z^{\prime}(v) + \sum _{h=1}^{L} \frac{d}{du}\phi (u,x_h) - \nonumber \\
&-& \int _{-\infty}^{+\infty}\frac {dv}{\pi}
\frac{d}{du}\phi (u,v)\frac {d}{dv}{\mbox {Im}}  \ln
[1+(-1)^{L} \, e^{iZ(v-i0^+)}]
\label {Zgen} \, .
\ea
We introduce the notations
\be
\sigma (u) = Z'(u) \, , \quad L'(u)= \frac {d}{du}{\mbox {Im}} \ln
[1+(-1)^{L} \, e^{iZ(u-i0^+)}] \label {not} \, ,
\ee
and pass to Fourier transforms $\hat f(k)=\int _{-\infty}^{+\infty} du \ e^{-iku} f(u)$, keeping in mind that
\ba
\hat \Phi _0(k)&=&- \frac{2\pi L e^{-\frac{|k|}{2}}}{ik} \, , \label {fou1} \\
\hat \Phi _H(k)&=& \frac{2\pi L}{ik} e ^{-\frac{|k|}{2}} [1-J_0(\sqrt{2}gk)] \, , \label {fou2} \\
\hat \phi _0(k)&=&\frac{2\pi e^{-|k|}}{ik} \, , \label {fou3} \\
\hat \phi _H(k,t)&=& -8i \pi ^2 \frac{e^{-\frac {|t|+|k|}{2}}}{k|t|}\Bigl [ \sum _{r=1}^{\infty} r (-1)^{r+1}J_r({\sqrt
{2}}gk) J_r({\sqrt {2}}gt)\frac {1-{\mbox {sgn}}(kt)}{2}
 + \nonumber \\
&+&{\mbox {sgn}} (t) \sum _{r=2}^{\infty}\sum _{\nu =0}^{\infty}
c_{r,r+1+2\nu}(g)(-1)^{r+\nu} \Bigl (
J_{r-1}({\sqrt {2}}gk) J_{r+2\nu}({\sqrt {2}}gt)-  \label {fou4}  \\
&-& J_{r-1}({\sqrt {2}}gt) J_{r+2\nu}({\sqrt {2}}gk)\Bigr ) \Bigr ] \, . \nonumber
\ea
We obtain the equation
\ba
\hat \sigma (k)&=& \frac{ik}{1-e^{-|k|}} \hat \Phi (k) -2\frac{ik e^{-|k|}}{1-e^{-|k|}} \hat {L}(k) +
\frac{ik}{1-e^{-|k|}} \int _{-\infty}^{+\infty}
\frac{dt}{4\pi ^2} \  \hat \phi _H (k,t) [\hat \sigma (t)-2 it \hat {L}(t) ] + \nonumber \\
&+& \frac{ik}{1-e^{-|k|}} \sum _{h=1}^L e^{ik x_h} \ \hat \phi _0 (k) + \frac{ik}{1-e^{-|k|}}\sum _{h=1}^L \int _{-\infty}^{+\infty}
\frac{dt}{2\pi }\ e^{it x_h} \ \hat \phi _H(k,t) \label {sigmak} \, ,
\ea
and for $\hat Z(k)$ the equation
\ba
\hat Z(k)&=& \frac{1}{1-e^{-|k|}} \hat \Phi (k) -2\frac{e^{-|k|}}{1-e^{-|k|}} \hat {L}(k) +
\frac{1}{1-e^{-|k|}} \int _{-\infty}^{+\infty}
\frac{dt}{4\pi ^2} \  \hat \phi _H (k,t) \ it \ [\hat Z (t)-2 \hat {L}(t) ] + \nonumber \\
&+& \frac{1}{1-e^{-|k|}} \sum _{h=1}^L e^{ik x_h} \ \hat \phi _0 (k) + \frac{1}{1-e^{-|k|}}\sum _{h=1}^L \int _{-\infty}^{+\infty}
\frac{dt}{2\pi }\ e^{it x_h} \ \hat \phi _H(k,t) \label {Zk} \, ,
\ea
which is the nonlinear integral equation for the counting function $Z(u)$, describing states of the $sl(2)$ sector.
We will find convenient to introduce the following function
\be
S(k)=\frac {\sinh \frac {|k|}{2}}{\pi |k|} \Bigl \{ \hat \sigma (k) + \frac{2ik e^{-|k|}}{1-e^{-|k|}}\hat L(k)+ \frac{\pi L}{\sinh \frac{|k|}{2}} \left (1-e^{-\frac{|k|}{2}} \right )
- \frac {2\pi e^{-|k|}}{1-e^{-|k|}}  \sum _{h=1}^{L} \left[ \cos kx_h- 1 \right] \Bigr \} \label {Sdef} \, ,
\ee
because, in Appendix A, we show that it satisfies the simple relation:
\be
\lim_{k \to 0} S(k) = \frac{\gamma  (g,L,s)}{2}  \, . \label {klip}
\ee
The function (\ref {Sdef}) satisfies the nonlinear equation
\ba
S(k)&=& \frac{L}{|k|} (1-J_0(\sqrt{2}gk) )+\frac{ik}{1-e^{-|k|}}\int _{-\infty}^{+\infty} \frac{dt}{2\pi }\hat \phi _H(k,t)
\left [ \frac{\sum \limits_{h=1}^L (\cos tx_h -1) +L(1-e^{-\frac{|t|}{2}})- \frac{it}{\pi} \hat L(t)}{1-e^{-|t|}} \right ] + \nonumber \\
&+&  \frac{ik}{1-e^{-|k|}}\int _{-\infty}^{+\infty} \frac{dt}{2\pi }\hat \phi _H (k,t) \frac{|t|}{2\sinh \frac{|t|}{2}} S(t) \, . \label {Seq0}
\ea
Now, the introduction of the 'magic kernel' \cite {BES}
\be
\hat K(t,t')=\frac{2}{tt'}\left [ \sum _{n=1}^{\infty}n J_n (t) J_n (t') + 2 \sum _{k=1}^{\infty} \sum _{l=0}^{\infty} (-1) ^{k+l}c_{2k+1,2l+2}(g) J_{2k}(t) J_{2l+1}(t') \right ] \, , \label {magic}
\ee
the use of the property, valid for $k>0$,
\be
\int _{-\infty}^{+\infty} dt \, \hat \phi _H(k,t) \, f(t) = 8i\pi ^2 g^2 \int _{0}^{+\infty} dt \, e ^{-\frac{t+k}{2}} \,
\hat K (\sqrt{2} gk, \sqrt{2} gt ) f(t) \, , \quad f (t)= f(-t) \, ,
\ee
and the restriction to $k>0$ allows to write the equation for $S(k)$ in the alternative way:
\ba
S(k)&=& \frac{L}{k} (1-J_0(\sqrt{2}gk) )-g^2 \int _{0}^{+\infty} \frac{dt}{\pi }e^{-\frac{t}{2}}
\hat K ( \sqrt{2} gk, \sqrt{2} gt ) \Bigl [ \frac{\pi t }{\sinh \frac{t}{2}} S(t) - \frac{2it}{1-e^{-t}}\hat {L}(t)+ \nonumber \\
&+& \frac{it}{1-e^{-t}} \hat \Phi _0 (t) + \left ( \frac{it}{1-e^{-t}} \hat \phi _0 (t) +2\pi \right ) \sum _{h=1}^L
e^{itx_h} \Bigr ] \, . \label {Seq}
\ea
Equations (\ref {Seq}, \ref {klip}) are our starting points for studying ABA contributions to anomalous dimension of twist operators.
As planned in the introduction, we will consider the minimal anomalous dimension state, go to the high spin limit and determine the predictions of ABA for the anomalous dimension up to orders $\frac{1}{s(\ln s)^n}$, $n\geq -1$. We therefore discuss in next section all the simplifications that equation
(\ref {Seq}) undergoes in the high spin limit.

\section{Ground state and high spin limit}
\setcounter{equation}{0}

In this section we start our study of the minimal anomalous dimension state.
For this state the positions of the internal holes are as close as possible to the origin, {\it i.e.} they satisfy the relations
\be
Z(x_h)=\pi (2h+1-L) \, , \quad h=1, \ldots , L-2 \, ,
\label{condhole}
\ee
while the positions of the two external holes are determined after solving the equations
\be
Z(x_{L-1})= -Z(x_L)=\pi (s+L-1) \, . \label{condexthole}
\ee
It follows that the positions of the Bethe roots $u_l$ are all greater in modulus than the positions of the internal holes, {\it i.e.} $|u_l|>x_h$, $h=1,...,L-2$. For our convenience we order Bethe roots $u_l$ in such a way that $u_l <u_{l'}$ if $l<l'$.

In the following we will find useful to integrate over the region in which Bethe roots are contained.
It is then very important to make the most convenient choice for the 'extrema' of integration, which naturally identify the points $\pm b$ which
separate the last/first root $u_{s}/u_{1}$ ($Z(u_{s}/u_{1})=\mp \pi (s+L-3)$) from the positive/negative external hole $x_L$/$x_{L-1}$:
we choose $b$ such that
\be
Z(\pm b)= \mp \pi (s+L-2) \, . \label{condb}
\ee
Then, we perform our analysis of the minimal anomalous dimension state in the high spin limit. We have to remark that
in this limit the set of operators (\ref {sl2op}) has been the object of an extensive activity 
\cite {ES,BES,BGK,FTT,BBKS1,BBKS2,BBKS3,BBKS4,CK1,CK2,BKK1,BKK2,FRS1,FRS2,FRS3,FRS4,FRS5,FRS6,BFR,FZ,FGR4,FIR}, also in perturbative QCD, see \cite {BDM1,BDM2,BDM3,BDM4,BDM5,BDM6}.
In the high spin limit, the position of the internal holes is proportional to $1/\ln s$, so it is very close to the origin: they will be determined by using (\ref {condhole}) in Section \ref {sec-4}.
On the other hand, in order to estimate the position of the two external or 'large' holes, we have to evaluate the counting function $Z(u)$ near the points $\pm b$, $b\sim s$, delimiting the interval in which Bethe roots reside. The result we will find, at their leading orders $O(s)$ and $O(s^0)$,
\be
x_{L}=-x_{L-1}=\frac{s}{\sqrt{2}}\left [1+\frac{L-1+\gamma  (g,L,s)}{2s}+ O\left (\frac{1}{s^{2}} \right ) \right ]
\label {extholes} \, ,
\ee
is proved in next subsection. We have to mention that the same formula (\ref {extholes}) was found for twist two in \cite {FZ}, by using results of \cite {BKM1,BKM2}. However, as far as we have understood, results of \cite {BKM1,BKM2} are proved only at one and two loops. Therefore, we would like to give a different and more general proof of (\ref {extholes}).

\subsection{Position of the external holes} \label {ehp}

When the spin is large, Bethe roots near the two 'extrema' $\pm b$ scale with $s$. In the proximity of $\pm b$, it is therefore convenient to rescale the variable $u$ of the counting function $Z(u)$: we will write $u=\bar u s$, where $\bar u$ will stay finite. Analogously, we will define $b=\bar b s$, with $\bar b$ finite.
From the definitions (\ref {Z}, \ref {Phi}, \ref {phi}) of the counting function, we have
\be
Z(\bar u s)=-2L \arctan 2\bar u s-2 \sum _{k=1}^{s} \arctan (\bar u s -\bar u_k s ) + \frac{\gamma (g,L,s) }{\bar u s} + O\left ( \frac{1}{s^2} \right ) \label {eq-in} \, .
\ee
We observe that the only 'higher loops' effect is in the last term, proportional to the anomalous dimension.
For $\bar u=\bar u_l$, where $\bar u_l s=u_l$ is a Bethe root, we expand the various functions for large $s$ and evaluate the sum over the Bethe
roots contained in (\ref {eq-in}) as an integral term plus an 'anomaly' \cite {Mat1,Mat2,BTZ1,BTZ2}. We obtain
\ba
Z(\bar u_l s)&=& -\pi L \textrm{sgn} (\bar u_l) + \frac{\gamma  (g,L,s) + L}{\bar u _l s}
+ 2 \int _{-\bar b}^{\bar b}d \bar v  \rho (\bar v) P \frac{1}{\bar u _l -\bar v} + \frac{\pi}{s} \rho '(\bar u_l) \coth \pi \rho (\bar u _l) - \nonumber \\
&-& \pi \sum  _{\stackrel {k=1} {k\not=l}}^s \textrm{sgn} (\bar u _l -\bar u_k) -\pi \sum _{h=1}^{L-2} \textrm{sgn} (\bar u_l -\bar x_h)  +2(L-2) \left [ \frac{\pi}{2} \textrm{sgn} (\bar u_l) - \frac{1}{\bar u_l s} \right ]+ O\left ( \frac{1}{s^2} \right ) \, , \label {first-eq}
\ea
where $\bar x_h=\frac{x_h}{s}$,
\be
\rho (\bar u)= - \frac{1}{2\pi s} \frac{d}{d\bar u} Z(\bar u s)
\ee
and where we used the relation \cite {Mat1,Mat2,BTZ1,BTZ2}
\ba
&& -2 \sum _{k=1}^s \arctan (u_l-u_k) -2 \sum _{h=1}^{L-2} \arctan (u_l-x_h)+\pi \sum _{\stackrel {k=1}
{k\not=l}}^s \textrm{sgn} (\bar u_l -\bar u_k) + \pi \sum _{h=1}^{L-2} \textrm{sgn} (\bar u _l -\bar x_h)= \nonumber \\
&=& \frac{1}{i} \sum _{\stackrel {k=1} {k\not=l}}^s  \ln \frac{u_l-u_k+i}{u_l-u_k-i}+ \frac{1}{i} \sum _{h=1}^{L-2} \ln \frac{u_l-x_h+i}{u_l-x_h-i}= \label {anomaly} \\
&=& 2 \int _{-\bar b}^{\bar b} d \bar v  \rho (\bar v) P \frac{1}{\bar u_l -\bar v} + \frac{\pi}{s} \rho '(\bar u_l) \coth \pi \rho (\bar u_l) + O\left ( \frac{1}{s^2} \right ) \, . \nonumber
\ea
We remark that, in order to obtain the last equality in (\ref {anomaly}) it is crucial to transform the sum into an integration from $-\bar b$ to $\bar b$, where $\bar b=b/s$ satisfies (\ref {condb}).
If, for instance, we transform the sum over Bethe roots and holes into an integration from the first $u_{1}=-u_s$ to the last $u_{s}$ root, we obtain an extra
$O(1/s)$ term $\frac{1}{u_l-u_{s}}+ \frac{1}{u_l+u_{s}}$ in the last line of  (\ref {anomaly}): in specific,
\ba
&& \frac{1}{i} \sum _{\stackrel {k=1} {k\not=l}}^s  \ln \frac{u_l-u_k+i}{u_l-u_k-i}+ \frac{1}{i} \sum _{h=1}^{L-2} \ln \frac{u_l-x_h+i}{u_l-x_h-i}= \label {anomaly2}  \\
&=& 2 \int _{-u_s}^{u_s} d \bar v  \rho (\bar v) P \frac{1}{\bar u_l -\bar v} + \frac{\pi}{s} \rho '(\bar u_l) \coth \pi \rho (\bar u_l) +
\frac{1}{s(\bar u_l-\bar u_{s})}+ \frac{1}{s(\bar u_l+\bar u_{s})}
O\left ( \frac{1}{s^2} \right ) \, . \nonumber
\ea
Sticking to formula (\ref {anomaly}), we remember that for the minimal anomalous dimension state and with our ordering of Bethe roots the value of the counting function on a generic root $u_l$ is given by the simple formula
\be
Z(\bar u _l s)=-\pi \sum _{\stackrel {k=1} {k\not=l}}^s  \textrm{sgn} (\bar u _l -\bar u_k) - \pi \sum _{h=1}^{L-2} \textrm{sgn} (\bar u _l -\bar x_h) \, . \label {Z-roots}
\ee
Property (\ref {Z-roots}) allows to simplify equation (\ref {first-eq}) as follows
\be
0=-2\pi \textrm{sgn} (\bar u_l) + \frac{4-L+\gamma  (g,L,s)}{\bar u_l s} +
2 \int _{-\bar b }^{\bar b}d \bar v  \rho (\bar v) P \frac{1}{\bar u_l -\bar v} + \frac{\pi}{s} \rho '(\bar u_l) \coth \pi \rho (\bar u_l) + O\left ( \frac{1}{s^2} \right )  \, . \label {eq-bar}
\ee
At the leading order, $O(s^0)$, we know that the equation to be satisfied, for all $\bar u$, is:
\be
0=- 2\pi \textrm{sgn} (\bar u)+2
\int _{-\bar b }^{\bar b}d \bar v  \rho (\bar v) P \frac{1}{\bar u  -\bar v}
\ee
whose solution is the well known \cite {KOR,ES} density
\be
\rho (\bar u)= \frac{1}{\pi} \ln \left ( \frac{\bar b + \sqrt{\bar b^2 -\bar u^2}}{\bar u} \right )^2 \, . \label {korden}
\ee
Using (\ref {korden}), we give an estimate of the last term in (\ref {eq-bar}),
\be
\frac{\pi}{s} \rho '(\bar u_l) \coth \pi \rho (\bar u_l)= \frac{1}{s} \left [ \frac{1}{2\bar b +2\bar u_l}-\frac{1}{2\bar b -2\bar u_l} - \frac{2}{\bar u_l} \right ] + O\left ( \frac{1}{s^2} \right ) \, , \label {cotanh}
\ee
which allows to find the function $\rho (\bar u)$ which satisfies (\ref {eq-bar}):
\be
\rho (\bar u)= \frac{1}{\pi} \ln \left ( \frac{\bar b + \sqrt{\bar b^2 -\bar u^2}}{\bar u} \right )^2 -
\frac{(2+\gamma  (g,L,s)-L) \delta (\bar u) + \delta (\bar u +\bar b) + \delta (\bar u -\bar b)}{2s}
+ O\left ( \frac{1}{s^2} \right ) \, . \label {sol-bar}
\ee
Using the form (\ref {sol-bar}) of the solution, we can determine the position of the extremum $\bar b$ through the relation
\be
\int _{-\bar b }^{\bar b} d\bar u \rho (\bar u)= - \frac{Z(b)-Z(-b)}{2\pi s}= 1+\frac{L-2}{s} \, ,
\ee
where we used (\ref {condb}), which gives
\be
\bar b= \frac{1}{2} \left (1+\frac{L-1+\gamma  (g,L,s)}{2s} \right )  + O\left ( \frac{1}{s^2} \right ) \, . \label {b-bar}
\ee
We remark that if we had transformed the sum over Bethe roots and holes into an integration from $u_1=-u_s$ to $u_s$ according to (\ref {anomaly2}), by
repeating all the steps until (\ref {b-bar}) we would have found the position of the largest root at leading and subleading order: this result is
\be
u_{s}=b + O(1/s) \, , \quad \label {us-b}
\ee
which, in particular, allows to give an estimate for $Z'(b)$:
\be
Z'(b) \sim \frac{Z(b)-Z(u_{s})}{b-u_{s}} \sim \frac{\pi}{O(1/s)} \sim O(s) \label {Zprimeb}
\ee
We will use this result for $Z'(b)$ in next subsection.

\medskip

We now pass to determine the position $x_L=\bar x_L s$, $\bar x_L>\bar b$, of the positive external hole.
We first compute equation (\ref {eq-in}) for $|\bar u|>|\bar b|$ (more precisely, $|\bar u| -|\bar b| =O(1)$):
\ba
Z(\bar u s)&=&-2L \arctan 2\bar u s-2 \sum _{k=1}^{s} \arctan (\bar u s -\bar u_k s ) + \frac{\gamma  (g,L,s)}{\bar u s} + O\left ( \frac{1}{s^2} \right )=
\nonumber \\
&=& -(L+s)\pi \textrm{sgn} (\bar u)+ \frac{L+\gamma}{\bar u s} + \frac{2}{s} \sum _{k=1}^{s} \frac{1}{\bar u-\bar u_k}  + O\left ( \frac{1}{s^2} \right )
\, . \label {eq-hol}
\ea
The sum over the Bethe roots is evaluated as
\ba
 \frac{2}{s} \sum _{k=1}^{s} \frac{1}{\bar u-\bar u_k}&=& 2 \int _{-\bar b}^{\bar b} d \bar v  \rho (\bar v) \frac{1}{\bar u -\bar v}- 2 \sum _{h=1}^{L-2} \frac{1}{\bar u s-x_h} +O(1/s^2) =  \nonumber \\
 &=& 2 \int _{-\bar b}^{\bar b} d \bar v  \rho (\bar v) \frac{1}{\bar u -\bar v} - \frac{2L-4}{\bar u s} + O(1/s^2) \, . \label {sum-roots}
\ea
We now insert (\ref {sol-bar}) into (\ref {sum-roots}) and use the result, valid for $|\bar u|>\bar b$:
\be
\int _{-\bar b}^{\bar b} \frac{d\bar v}{\bar u-\bar v} \ln \left ( \frac{\bar b + \sqrt{\bar b^2 -\bar v^2}}{\bar v} \right )^2=
i\pi \ln \frac{i\bar u \sqrt{1-\frac{\bar b^2}{\bar u^2}}+\bar b}{i\bar u \sqrt{1-\frac{\bar b^2}{\bar u^2}}-\bar b}
\, .  \label {rho-integ}
\ee
Inserting the resulting expression for (\ref {sum-roots}) into (\ref {eq-hol}), we eventually arrive at the formula
\be
Z(\bar u s)=-(L+s)\pi \textrm{sgn} (\bar u) + \frac{2}{\bar u s}-\frac{1}{2s}\left ( \frac{1}{\bar u +\bar b}+ \frac{1}{\bar u -\bar b} \right )+2i
\ln \frac{i\bar u \sqrt{1-\frac{\bar b^2}{\bar u^2}}+\bar b}{i\bar u \sqrt{1-\frac{\bar b^2}{\bar u^2}}-\bar b}
+ O\left ( \frac{1}{s^2} \right ) \, , \label {320}
\ee
which is certainly valid for $|\bar u|-|\bar b| =O(1)$.
Now, the position of the (positive) external holes is fixed by the condition $Z(\bar x_L s)=-(L+s)\pi +\pi$. Imposing that on (\ref {320}), we find
\be
\bar x_L=\sqrt{2}\bar b + O(1/s^2)  \Rightarrow \bar x_L=\frac{1}{\sqrt{2}} \left (1+\frac{L-1+\gamma (g,L,s)}{2s} \right )  + O\left ( \frac{1}{s^2} \right ) \, .
\ee
We observe that such result agrees when $L=2$ with the zeroes of the transfer matrix which one can obtain from
expressions contained in \cite {BKM1,BKM2}. This is an important check for our findings.

\medskip

\subsection{High spin limit of nonlinear terms} \label {hispinnl}

Another important simplification occurring for large spin concerns the nonlinear term (containing $\hat L'(t)$) which appears in (\ref {Seq}).
In this subsection we extend to all loops the result of \cite {FRS1,FRS2,FRS3,FRS4,FRS5,FRS6}. Some of the results of this section have been already announced, but not completely proved, in \cite {FR}. We will fill that gap here: at the end, we are able to show that
\be
g^2 \int_0^{+\infty} \frac{dt}{\pi} e^{-\frac{t}{2}} \hat K(\sqrt{2} g k,\sqrt{2} g t)
\frac{2 i t}{1-e^{-t}} \hat L(t) =2g^2 \ln 2 \hat K(\sqrt{2} g k,0) + O\left(\frac{1}{s^2}\right) \, . \label  {nlt}
\ee
This means that, in our approximation, nonlinearity effects in equation (\ref {Seq}) are under control.

In our equation (\ref {Seq}) nonlinearity appears in the following integral:
\be
\hat {NL} (k) =g^2 \int_0^{+\infty} \frac{dt}{\pi} e^{-\frac{t}{2}} \hat K(\sqrt{2} g k,\sqrt{2} g t)\frac{2 i t}{1-e^{-t}} \hat L(t) \; . \label {b1}
\ee
It is convenient to pass to the coordinate space and to define
\ba \label{Iu}
I^{\alpha }(u)&=& - 2 \int_{0}^{+\infty} \frac{dt}{2\pi} \cos tu  \frac{2ite^{-\alpha t}}{1-e^{-t}} \hat L (t)\nonumber\\
    &=& \int_{-\infty}^{+\infty} \frac{dv}{i\pi}\left[ \psi^\prime (\alpha -iu+iv) -\psi^\prime (\alpha +iu-iv)\right] L(v)\; . \label {I-alfa-def}
\ea
We can keep $\alpha $ generic, having in mind that the case $\alpha = \frac{1}{2}$ is relevant for our case (\ref {b1}):
\be
NL(u)=2 \int _{0}^{+\infty} \frac{dk}{2\pi} \cos ku \ \hat {NL} (k) = - \int _{0}^{+\infty} dv K\left ( \frac{u}{\sqrt{2}g}, \frac{v}{\sqrt{2}g} \right ) I ^{\frac{1}{2}}(v) \, ,
\ee
where
\be
K\left ( \frac{u}{\sqrt{2}g}, \frac{v}{\sqrt{2}g} \right )=8g^2  \int _{0}^{+\infty} \frac{dk}{2\pi} \int _{0}^{+\infty} \frac{dt}{2\pi} \cos ku \cos tv \hat K(\sqrt{2} gk,\sqrt{2} gt) \, .
\ee
In general we split $I^{\alpha }(u)$ as $I^{\alpha }(u)=I^{\alpha }_{in}(u) + I^{\alpha }_{out}(u)$, where
\ba
I^{\alpha }_{in}(u)&=&\int_{-b}^{b} \frac{dv}{i\pi}\left[ \psi^\prime (\alpha -iu+iv) -\psi^\prime (\alpha +iu-iv)\right] L(v) \, ,  \\
I^{\alpha }_{out}(u)&=&\int_{|v|>b} \frac{dv}{i\pi}\left[ \psi^\prime (\alpha -iu+iv) -\psi^\prime (\alpha +iu-iv)\right] L(v) \, .
\ea
Then, $I^{\alpha }_{in}(u)$ is evaluated using formula (2.17) of \cite {rep}:
\ba
I^{\alpha }_{in}(u)&=& - i B_{2}(1/2) \frac{\psi ' (\alpha -iu+ib) -\psi '(\alpha+iu-ib)-\psi ' (\alpha -iu-ib) +\psi '(\alpha+iu+ib)}{Z'(b)}+ \nonumber \\
&+&  O(1/Z'(b)^3) = \frac{2 B_{2}(1/2)}{Z'(b)} \left [ \frac{u-b}{\alpha ^2+(u-b)^2}-   \frac{u+b}{\alpha ^2+(u+b)^2} \right ] +  O(1/Z'(b)^3) \, .
\ea
Now, we remember that $Z'(b)=O(b)$ (see \ref {Zprimeb}); in addition, in the high spin limit we are allowed to consider $u \ll s$. Therefore, we conclude that $I^{\alpha }_{in}(u)=O(1/s^2)$ and, consequently,
\be
I^{\alpha }(u)=\int_{|v|>b} \frac{dv}{i\pi}\left[ \psi^\prime (\alpha -iu+iv) -\psi^\prime (\alpha +iu-iv)\right] L(v) + O\left ( \frac{1}{s^2} \right ) \, . \label {Ialfau}
\ee
Since we can restrict $I^{\alpha }(u)$ to $|u|\ll s$, we develop the $\psi $ functions in the integrand for large $v$. We obtain
\be
I^{\alpha }(u)=-\frac{4}{\pi}\int_{b}^{+\infty} \frac{dv}{v} L(v) + O\left ( \frac{1}{s^2} \right ) \, , \quad |u|\ll s \, .
\ee
Integrating by parts we can write down
\be
I^{\alpha }(u)=\frac{4}{\pi} \ln b \ L(b)+\frac{4}{\pi}\int_{b}^{+\infty} dv \ln v \ L'(v) + O\left ( \frac{1}{s^2} \right ) \, , \quad |u|\ll s \, . \label {Iu2}
\ee
We then use the fact that $L(b)=0$ and the identity
\be
\ln x_L= -\int _{b}^{+\infty} \frac{dv}{2\pi} \ln v \ Z'(v) +  \int _{b}^{+\infty} \frac{dv}{\pi} \ln v \ L'(v)
\ee
to obtain
\be
I^{\alpha }(u)=4 \ln x_L + \frac{2}{\pi}\int _{b}^{+\infty} dv \ln v \ Z'(v) \label {Iu3} \, .
\ee
In order to perform the integration in (\ref {Iu3}), we need an estimate of $Z'(v)$ when $v>b$.
In Appendix \ref {Zprime} we prove that
\be
Z'(v)=-\frac{4b}{v} \frac{1}{\sqrt{v^2-b^2}} + O\left ( \frac{1}{b^3} \right ) \, , \quad v>b \, .
\ee
Integration in (\ref {Iu3}) is then performed exactly
\be
-\frac{8b}{\pi}\int _{b}^{+\infty} dv \frac{\ln v}{v} \frac{1}{\sqrt{v^2-b^2}}= - 4 \ln b - 4 \ln 2   \label {Iu4} \, .
\ee
Plugging (\ref {Iu4}) into (\ref {Iu3}) and using the equality $x_L=\sqrt{2} b+O(1/s)$, we obtain
\be
I^{\alpha }(u)=-2\ln 2  + O\left(\frac{1}{s^2}\right) \, , \quad |u|\ll s \, . \label  {nlt0}
\ee
Passing now to the kernel $K$, its evaluation in the coordinate space shows the following behaviour
\ba
K \left ( \frac{u}{\sqrt{2}g}, \frac{v}{\sqrt{2}g} \right )&=& - \frac{1}{\pi ^2} \ln \left [ 1- \frac{g^4}{4x(u)^2x(v)^2} \right ] \, , \quad |u|,|v| \geq \sqrt{2}g \, , \nonumber \\
&& \label {Ku}\\
K \left ( \frac{u}{\sqrt{2}g}, \frac{v}{\sqrt{2}g} \right ) &=& - \frac{1}{2\pi ^2 } \ln \left ( \left [ 1+ \frac{g^2 e^{2i \arcsin \frac{u}{\sqrt{2}g}}}{2x(v)^2} \right ]  \left [ 1+ \frac{g^2 e^{-2i \arcsin \frac{u}{\sqrt{2}g}}}{2x(v)^2} \right ] \right )
 \, , \quad |u|\leq \sqrt{2}g \, , |v| \geq \sqrt{2}g \nonumber \, .
\ea
Therefore,
\ba
NL(u)&=&-  \int _{0}^{+\infty} dv K\left ( \frac{u}{\sqrt{2}g}, \frac{v}{\sqrt{2}g} \right ) I ^{\frac{1}{2}}(v) = \\
&=& 2 \ln 2 \int _{0}^{\Lambda} dv  K\left ( \frac{u}{\sqrt{2}g}, \frac{v}{\sqrt{2}g} \right )
- \int _{\Lambda}^{+\infty} dv K\left ( \frac{u}{\sqrt{2}g}, \frac{v}{\sqrt{2}g} \right ) I ^{\frac{1}{2}}(v) + O \left(\frac{1}{s^2}\right) \, , \nonumber
\ea
where $\Lambda \sim s $ is a cutoff such that for $ u < \Lambda $ approximation (\ref {nlt0}) can be used. When $s\rightarrow +\infty$, in the first integral we replace $\Lambda $ with $+\infty$;
in the second integral we estimate $K$ using (\ref {Ku}) and $I^{\frac{1}{2}}(v)$ by means of (\ref {I-alfa-def}),
using that $\psi '(z) \sim 1/z$ for large $z$:
\be
v \rightarrow +\infty \quad \Rightarrow \quad K\left ( \frac{u}{\sqrt{2}g}, \frac{v}{\sqrt{2}g} \right ) \sim \frac{1}{v^2} \, , \quad I^{\frac{1}{2}}(v) \sim \frac{1}{v} \, ,
\ee
which therefore imply that
\be
\Lambda \rightarrow \infty \quad \Rightarrow \quad \int _{\Lambda}^{+\infty} dv K\left ( \frac{u}{\sqrt{2}g}, \frac{v}{\sqrt{2}g} \right ) I ^{\frac{1}{2}}(v) \sim \frac{1}{\Lambda ^2} \sim \frac{1}{s^2} \, .
\ee
Putting all together we find out
\be
\hat {NL}(k)=2g^2 \ln 2 \hat K(\sqrt{2} g k,0) + O\left(\frac{1}{s^2}\right) \, . \label  {nlt1}
\ee

\section{High spin results from ABA: up to order $\frac{1}{s}$} \label {sec-4}
\setcounter{equation}{0}

Having analysed all the simplifications occurring in the high spin limit,
let us come back to equation (\ref {Seq}). We insert formula (\ref {extholes}) for the position of the external holes, use relation
(\ref {nlt}) for the non linear term and work out all the 'known' terms. We end up with the following integral equation:
 \begin{eqnarray}
&&S(k) = 4g^2 \ln s \hat K( \sqrt{2}gk, 0) + 4g^2 \int _{0 }^{+\infty} \frac{dt}{e^{t}-1} \hat K^{*} (\sqrt{2}gk,\sqrt{2}gt)+ \frac{2g^2}{s}(L+\gamma (g,L,s)-1) \hat K( \sqrt{2}gk, 0) + \nonumber  \\
&+&  \frac {L}{k}[1-J_0({\sqrt {2}}gk)]+4g^2 \gamma _E \hat K( \sqrt{2}gk, 0) + g^2 (L-2) \int _{0 }^{+\infty}dt  e^{-\frac{t}{2}}\hat K (\sqrt{2}gk,\sqrt{2}gt)  \frac {1-e^{\frac {t}{2}}}{\sinh \frac {t}{2}}  - \label {Sdef2} \\
&-& g^2 \int _{0 }^{+\infty} {dt}
\hat K (\sqrt{2}gk,\sqrt{2}gt)
 \frac {\sum _{h=1}^{L-2} \left[ \cos t x_h -1 \right]}{ \sinh \frac {t}{2}}-
g^2 \int _{0 }^{+\infty}{dt}
e^{-\frac{t}{2}} \hat K (\sqrt{2}gk,\sqrt{2}gt)
  \frac { t}{\sinh \frac {t}{2}}S(t)  + O\left ( s^{-1}(\ln s)^{-\infty} \right )
\, , \nonumber
\end{eqnarray}
where $\hat K^{*} (t,t')=\hat K(t,t')-\hat K(t,0)$.
The particular form of the known terms in (\ref {Sdef2}), together with condition (\ref {condhole}) for the internal holes, suggests that $S(k)$ expands in (inverse) powers \footnote {With $O\left ( s^{-1}(\ln s)^{-\infty} \right )$
we denote terms going to zero faster than $\frac{1}{s}$ times any inverse power of $\ln s$.} of $\ln s$,
\be
S(k)=\sum _{n=-1}^{\infty}\frac{S^{(n)}(k)}{(\ln s )^{n}}+ \sum _{n=-1}^{\infty}\frac{\check S ^{(n)}(k)}{s(\ln s )^{n}}+ O\left ( s^{-1}(\ln s)^{-\infty} \right ) \label{Shigh} \, .
\ee
And, consistently with (\ref {Shigh}), the condition for the internal holes (\ref {condhole}) is solved by the following Ansatz on their positions:
\be
x_h = \sum_{n=1}^\infty \left ( \alpha_{n,h} + \frac{\check \alpha_{n,h}}{s} \right ) (\ln s)^{-n} + O\left ( s^{-1}(\ln s)^{-\infty} \right ) \, . \label {upos}
\ee
For the anomalous dimension $\gamma (g,L,s)=2S(0)$, therefore, we have the expansion,
\ba \label{gamma}
\gamma (g,L,s) &=& f(g) \ln s + f_{sl}(g,L)+ \sum _{n=1}^{\infty}\frac{\gamma ^{(n)}(g,L)}{(\ln s )^{n}}+ \nonumber \\
&+& \sum _{n=-1}^{\infty}\frac{\check \gamma ^{(n)}(g,L)}{s(\ln s )^{n}}+ O\left ( s^{-1}(\ln s)^{-\infty} \right ) \, ,
\ea
where the scaling functions $f(g)$, $f_{sl}(g,L)$ appear also in other contexts: for instance, $f(g)$ is twice the cusp anomalous dimension of Wilson loops \cite {KM1,KM2}.

For our purposes, it is important to remark that the strong coupling limits of $f(g)$ \cite {BKK1,BKK2} and $f_{sl}(g,L)$ \cite {FZ,FGR4} agree with string theory computations.
This shows that such functions are actually wrapping independent and, consequently, the anomalous dimension is wrapping independent at its leading orders $\ln s$ and $(\ln s)^0$.

\medskip

After these first considerations, we come back to (\ref {Sdef2}): the 'known' terms driving the equations for $S^{(-1)}(k)$ and $S^{(0)}(k)$ are contained in the first two lines of (\ref {Sdef2}). The structure of such driving terms implies the following equalities between densities
\be
\check S^{(-1)} (k) = \frac{f(g)}{2} S^{(-1)} (k) \, , \quad
\check S^{(0)} (k) = \frac{f_{sl} (g,L)+ L-1}{2}  S^{(-1)} (k) \, , \label {S-funct}
\ee
which translate in terms of anomalous dimensions to the equalities \cite {FZ,BFTT}
\be
\check \gamma ^{(-1)}(g,L)=\frac{1}{2} [f(g)]^2 \, , \quad \check \gamma ^{(0)}(g,L)=\frac{1}{2}f(g)[L-1+
f_{sl}(g,L)] \label {checkgamma} \, .
\ee

It is possible to obtain analogous relations for $\check \gamma^{(n)}(g,L)$ expressed in terms of the $\gamma^{(n)}(g,L)$, for $n\geq 1$. The first step is a standard procedure for integral equations with a separable kernel, called Neumann expansion \cite{KL}, applied to the equation (\ref{Sdef2}) for $S(k)$:
\ba
&& S(k)=\sum _{p=1}^{\infty} \mathcal{S}_p (g,L,s) \frac {J_p({\sqrt 2}gk)}{k} \, , \quad
\mathcal{S}_p (g,L,s)=\sum _{n=-1}^{\infty} \left [ S_p^{(n)}(g,L)+\frac{1}{s} \check S_p^{(n)}(g,L) + O(1/s^2) \right ] (\ln s )^{-n}
\Rightarrow \nonumber \\
&& \gamma^{(n)}(g,L) = {\sqrt 2} g S_1^{(n)}(g,L) \, , \ \ \check \gamma^{(n)}(g,L) = {\sqrt 2} g \check S_1^{(n)}(g,L) \, .
\ea
This procedure is fully explained in this application in Appendix \ref{neu_exp}. For $n=1,2,3,4,5$ we obtain
\ba
\check S^{(1)}_p (g,L) &=& 0 \nonumber \\
\check S^{(2)}_p (g,L) &=& 2\pi  \tilde S^{(1)}_{p}(g) \sum\limits_{h=1}^{L-2} \alpha _{1,h}
\check \alpha _{1,h}+ \frac{S_p^{(-1)}(g)}{2}  \gamma^{(2)} (g,L) \nonumber \, , \\
 \check S^{(3)}_p (g,L) &=& 2\pi  \tilde S^{(1)}_{p}(g) \sum\limits_{h=1}^{L-2} \left ( \alpha _{2,h} \check \alpha _{1,h}+ \alpha _{1,h} \check \alpha _{2,h}\right )+ \frac{S_p^{(-1)}(g)}{2}  \gamma^{(3)} (g,L) \, ,
\label {gamma 234}
\\ \check S^{(4)}_p (g,L) &=& 2\pi  \tilde S^{(1)}_{p}(g) \sum\limits_{h=1}^{L-2} \left ( \alpha _{1,h} \check \alpha _{3,h}+ \alpha _{2,h} \check \alpha _{2,h}+ \alpha _{3,h} \check \alpha _{1,h}\right )
- \frac{\pi}{3} \tilde S^{(2)}_{p}(g)\sum\limits_{h=1}^{L-2} (\alpha _{1,h})^3 \check \alpha _{1,h}  + \frac{S_p^{(-1)}(g)}{2}  \gamma^{(4)} (g,L) \nonumber \, ,
\\ \check S^{(5)}_p (g,L) &=& 2\pi  \tilde S^{(1)}_{p}(g) \sum\limits_{h=1}^{L-2} \left(\alpha_{1,h} \check \alpha_{4,h}+\alpha_{2,h}\check \alpha_{3,h} + \alpha_{3,h}\check \alpha_{2,h}+\alpha_{4,h}\check \alpha_{1,h}\right)\nonumber
\\ && - \frac{\pi}{3} \tilde S^{(2)}_{p}(g)\sum\limits_{h=1}^{L-2} \left( 3(\alpha_{1,h})^2 \alpha_{2,h}\check \alpha_{1,h} +(\alpha_{1,h})^3 \check \alpha_{2,h}\right)+\frac{S_p^{(-1)}(g)}{2}  \gamma^{(5)} (g,L) \, , \nonumber
\ea
where $\tilde S^{(1)}_{p}(g)$ and $\tilde S^{(2)}_p(g)$ belong to a set of 'reduced coefficients', satisfying the system (\ref{tildeSsystem}), reported also in Appendix \ref{neu_exp}.

These expressions are still quite involved, but they can be significantly simplified. This can be done through the following steps.
\begin{itemize}
\item After introducing the notation
\begin{eqnarray} \label{defsigmarn}
\frac{d^{r}}{d u^{r}} \sigma(u=0)& = \sum\limits _{n=-1}^\infty\left(\sigma_r^{(n)}+\frac{\check \sigma^{(n)}_{r}}{s}
+O\left(\frac{1}{s^2}\right)\right) \, (\ln s)^{-n} \, ,
\end{eqnarray}
we 'invert' relation (\ref{condhole}), expressing $\alpha_{m,h}$ and $\check \alpha_{m,h}$ in terms of the densities and their derivatives in zero, {\it i.e.} in terms of the coefficients $\sigma_r^{(n)}$ and $\check \sigma_r^{(n)}$.
In performing this procedure we use techniques and results of \cite{FGR5}. Then, we plug the obtained expressions for $\alpha_{m,h}$, $\check \alpha_{m,h}$ in (\ref{gamma 234}). Detailed calculations are shown in Appendix \ref{alpha_den}, where we have also listed the full expressions for the first $\check S_p^{(n)}(g,L)$ (relations (\ref{gamma 234 sigma})).
\item  Then, we use the following relations, proven in Appendix \ref {sigf}:
\be \label{relfg}
f(g)=2\frac{\check \sigma_0^{(-1)}}{\sigma_0^{(-1)}}=2\frac{\check \sigma_2^{(-1)}}{\sigma_2^{(-1)}} \, ,
\ee
\be \label{relfslg}
f_{sl} (g,L)=2 \frac{\check \sigma_0^{(0)}}{\sigma_0^{(-1)}} - (L-1) \, ,
\ee
\be \label{relgamma2}
\gamma^{(2)}(g,L)= 2\frac{\check \sigma_0^{(2)}}{\sigma_0^{(-1)}}+4\frac{\sigma_0^{(2)} \check \sigma_0^{(-1)}}{{(\sigma_0^{(-1)})}^2}\, .
\ee
With the help of these formul{\ae} it is possible to compare the complicated relations (\ref{gamma 234 sigma}) with analogous results for $S_p^{(n)}(g,L)$, found in \cite {FGR5} and reported in Appendix \ref {gamman}, ending up with the following simple and compact expressions:
\ba
\check S^{(1)}_p (g,L)&=&0 \label{Srec1} \, , \\
\check S^{(2)}_p (g,L)&=&\frac{S_p^{(-1)}(g)}{2}\gamma ^{(2)}(g,L)-f(g)S_p^{(2)} (g,L)\label{Srec2} \, , \\
\check S^{(3)}_p (g,L)&=&\frac{S_p^{(-1)}(g)}{2}\gamma^{(3)} (g,L) -(f_{sl} (g,L)+L-1)S_p^{(2)}(g,L) -\frac{3}{2} S_p^{(3)}(g,L) f(g) \label{Srec3} \, , \\
\check S^{(4)}_p(g,L)&=&\frac{S_p^{(-1)}(g)\gamma ^{(4)}(g,L)}{2}-2S_p^{(4)}(g,L)f(g)-\frac{3}{2} S_p^{(3)}(g,L) (f_{sl} (g,L)+L-1) \label{Srec4} \, , \\
\check S^{(5)}_p (g,L)&=&\frac{S_p^{(-1)}(g)}{2}\gamma ^{(5)}(g,L)-\frac{5}{2}S_p^{(5)} (g,L)f(g)-2 S_p^{(4)} (g,L)(f_{sl} (g,L)+L-1)\label{Srec5}\\ &&-S_p^{(2)} (g,L)\gamma^{(2)} (g,L)\nonumber \, .
\ea
For anomalous dimensions, such relations read
\ba
\check \gamma^{(1)} (g,L)&=& 0 \label{1rec1} \, , \\
\check \gamma^{(2)} (g,L)&=&-\frac{f(g)}{2}\gamma^{(2)} (g,L)\label{1rec2} \, , \\
\check \gamma^{(3)} (g,L)&=&-f(g)\gamma^{(3)} (g,L) -(f_{sl} (g,L)+L-1)\gamma^{(2)} (g,L)\label{1rec3} \, , \\
\check \gamma^{(4)} (g,L)&=&-\frac{3}{2} f(g) \gamma^{(4)} (g,L) -\frac{3}{2} (f_{sl} (g,L)+L-1)\gamma^{(3)} (g,L)\label{1rec4} \, , \\
\check \gamma^{(5)}(g,L)&=&-2f(g)\gamma^{(5)} (g,L) -2(f_{sl} (g,L)+L-1)\gamma^{(4)} (g,L)-{\left(\gamma^{(2)}(g,L)\right)}^2\label{1rec5}\, .
\ea
\end{itemize}
Relations (\ref {checkgamma}) and (\ref {1rec1}-\ref {1rec5}) are the prediction of ABA for anomalous dimensions at various orders $\frac{1 }{s(\ln s)^{n}}$, $n=-1,...5$. In next section we will show that they agree with predictions coming from self-tuning and reciprocity relations.

\section{$\frac{1}{s}$ contributions from functional relations}
\setcounter{equation}{0}

Self-tuning and reciprocity relations were summarised in formul{\ae} (\ref {self-tun}, \ref {reci}, \ref {spinC}). We remember notations (\ref {gamma}) for the high spin expansion of the anomalous dimension:
\ba \label{gamma2}
\gamma(g,L,s) &=& f(g) \ln s + f_{sl}(g,L)+ \sum _{n=1}^{\infty}\frac{\gamma ^{(n)}(g,L)}{(\ln s )^{n}}+ \nonumber \\
&+& \sum _{n=-1}^{\infty}\frac{\check \gamma ^{(n)}(g,L)}{s(\ln s )^{n}}+ O\left ( s^{-1}(\ln s)^{-\infty} \right ) \, .
\ea
Comparing (\ref {gamma2}) with (\ref {self-tun}, \ref {reci}, \ref {spinC}), we obtain that the leading terms of ${P}(s)$ should read
\be
{P}(s)=f(g) \ln C (s) + f_{sl}(g,L)+ \sum _{n=1}^{\infty} \frac{\gamma ^{(n)} (g,L)}{(\ln C(s))^{n}}+ O (1/C^2) \, . \label {highspin}
\ee
Developing $C(s)$ in the same regime
\be
C(s)^2=\left (s+\frac{L}{2}-1 \right) \left (s+\frac{L}{2}\right )\Rightarrow C (s)= s+\frac{L-1}{2}+O(1/s) \, , \label {highspinC2}
\ee
and putting together (\ref {self-tun}, \ref {highspin}, \ref {highspinC2}),
we end up with the following prediction for the anomalous dimension
\ba \label{rec}
\gamma (g,L,s)&=& f(g) \ln s + f_{sl}(g,L)+ \sum _{n=1}^{\infty} \frac{\gamma ^{(n)} (g,L)}{(\ln s)^{n}}  \nonumber \\
&+& \frac{\ln s}{2s} [f(g)]^2+ \frac{1}{2s}f(g)(L-1+f_{sl}(g,L))+ \frac{f(g)}{2s} \sum _{n=1}^{\infty}
\frac{\gamma ^{(n)}(g,L)}{(\ln s)^n}- \label {recS} \\
&-& \sum _{n=1}^{\infty}
n \frac{\gamma ^{(n)}(g,L)}{2s(\ln s)^{n+1}} \left [ f(g) \ln s + f_{sl}(g,L)+L-1+ \sum _{m=1}^{\infty}\frac{\gamma ^{(m)}(g,L)}{(\ln s )^{m}} \right ] + O\left ( s^{-1}(\ln s)^{-\infty} \right ) \, . \nonumber
\ea
Working out this formula for orders $\frac{1}{s (\ln s )^n}$, $n=-1,...,5$, we find formul{\ae} which coincide with (\ref {checkgamma}), for $n=-1,0$ and with (\ref {1rec1}, \ldots , \ref {1rec5}), for $n=1,...,5$. Therefore, our findings in Section 4 agree with self-tuning and reciprocity predictions.
We would like to stress the fact that this conclusion holds for all values of the coupling $g$, {\it i.e.} it is a nonperturbative statement on the high spin expansion of (asymptotic) anomalous dimension.

\medskip

\textbf{Remark 1}: Formul{\ae} (\ref {S-funct}) and (\ref {Srec1},...,\ref {Srec5}) seem to indicate that more generally functional relations similar to (\ref {self-tun}, \ref {reci}) should hold for (the high spin expansion of) the function $S(k)$.
On the basis of our results, we are naturally led to make the following proposal for a self-tuning relation involving the
coefficients of the Neumann expansion of the function $S(k)$,
\be
\mathcal{S}_p (g,s,L)=\mathcal{P}_p \left (s+\frac{1}{2}\gamma (g,s,L)\right ) \, , \label {Sreci2}
\ee
where $\mathcal{P}_p(s)$ satisfies an high spin expansion analogous to (\ref {reci}):
\be
\mathcal{P}_p(s)=\sum _{n=0}^{\infty} \frac{a^{(n)}_p(\ln C(s))}{C(s)^{2n}} \, , \label {reci2}
\ee
where $C(s)$ is given by (\ref {spinC}). In particular, formula (\ref {Sreci2}) has the advantage to furnish immediately the self-tuning (and reciprocity) relations for all the higher conserved charges \cite {BM}
\be
Q_r (g,L,s)= P_r \left (s+\frac{1}{2}\gamma (g,s,L)\right ) \, .
\ee
Furthermore, we may suppose that, because of integrability, the statement above is equivalent to the self-tuning (and reciprocity) of the counting function (\ref{Sreci2}).

\medskip

\textbf{Remark 2}: Concerning the leading terms in the high spin expansion (\ref {gamma}) of the minimal anomalous dimension, we already commented that, since the strong coupling limit of $f(g)$ and $f_{sl}(g,L)$ agrees with string theory calculations (and many gauge loop calculations), there are good reasons to believe that anomalous dimension
at the orders $\ln s$ and $(\ln s)^0$ is free from wrapping contributions. Then, if we suppose that self-tuning and reciprocity are exact symmetries, from (\ref {checkgamma}, \ref {recS}) it follows that also $\check \gamma ^{(-1)}$ and $\check \gamma ^{(0)}$ are wrapping-free. Then, one could expect that all the terms that are in between $f_{sl}(g,L)$ and $\check \gamma ^{(-1)}$ (the various $\gamma ^{(n)}(g,L)$) are also not affected by wrapping. If we suppose this, use again self-tuning and reciprocity and compare (\ref {1rec1}-\ref {1rec5}) with (\ref {recS}), we are able to conclude that all the functions $\check \gamma ^{(n)}(g,L)$ do not depend on wrapping either.

Even if are aware that our arguments do not provide a proof of the fact that at high spin wrapping
diagrams start contributing at orders $\frac{(\ln s)^2}{s^2}$, we however think that our results
provide some nonperturbative hints of this property.

\medskip

\textbf{Remark 3}: In the previous remark we gave some evidence in favour of the fact that when $s\rightarrow +\infty$ and the twist
$L$ is fixed, for all values of the coupling constant wrapping diagrams start contributing at the order $\frac{(\ln s)^2}{s^2}$.
We would like to stress that this conclusion depends on the particular order in which limits are performed: indeed, we first sent $s\rightarrow +\infty$, with $L$ and $g$ fixed,
then, possibly, we could have made the limit $g\rightarrow +\infty$. Obviously, the our situation is different from what happens in the peculiar regime of semiclassical strings: in this case, indeed, $s$ and $g$ go together to infinity, with their ratio $\mathcal{S}=\frac{s}{2\sqrt{2}\pi g}$ kept fixed, and wrapping contribution enter already at the order $\frac{1}{\ln \mathcal{S}}$ \cite {GSSV1,GSSV2,GSSV3,GSSV4} .

\section{Conclusions}
\setcounter{equation}{0}

We studied the high spin limit of twist operators in the $sl(2)$ sector of ${\cal N}=4$ SYM.
Using ABA equations rewritten as one NLIE, we computed the minimal anomalous dimension
up to orders $\frac{1}{s(\ln s)^{n}}$ -- in detail: Section 4 and formul{\ae} (\ref {1rec1},...,\ref {1rec5}) --, proving eventually that our results satisfy (for all values of the coupling $g$) the self-tuning and reciprocity properties (\ref {self-tun}, \ref {reci}, \ref {spinC}). As a consequence, in Remark 2 above we could give some clues supporting the idea that in the high spin limit wrapping corrections start contributing at order $\frac{(\ln s)^2}{s^2}$ for any twist $L$.

In addition, as a byproduct of our analysis we provided also the following new results:

1) Exact connection between a nonlinear function of the counting function and the (asymptotic) anomalous dimension (formula (\ref {klip}) and Appendix A).

2) Evaluation of the external holes position at the (subleading) order $s^0$ (Section 3.1).

3) Proposal for self-tuning and reciprocity relations satisfied by
the function (directly related to the counting function) $S(k)$
(Remark 1 of Section 5 and formul{\ae} (\ref {Sreci2}, \ref {reci2}).

Eventually, we are confident that further analysis of the last issue may shed light on how to construct, order by order in $s$, an 'effective' (non)linear integral equation which takes into account the wrapping corrections as well.

\medskip

\medskip

{\bf Acknowledgements} We thank gratefully B. Basso for scientific discussions. As for travel financial support, INFN IS Grant GAST, the UniTo-SanPaolo research  grant Nr TO-Call3-2012-0088, the ESF Network {\it HoloGrav} (09-RNP-092 (PESC)) and the MPNS--COST Action MP1210 are kindly acknowledged. G.I. acknowledges E.U., Italian Republic and Calabria Region for funding through Regional Operative Program (ROP) Calabria ESF 2007/2013 - IV Axis Human Capital - Operative Objective M2.

\appendix

\section{Connection between the counting function and (asymptotic) anomalous dimension} \label {kotlipgen}
\setcounter{equation}{0}

It is convenient to rewrite equation (\ref {Seq0}) in the following form:
\ba
S(k)&=& \frac{L}{|k|}\left [ 1-J_0(\sqrt{2}gk) \right ] +\frac{ik}{2\pi |k|} e^{\frac{|k|}{2}}\sum _{h=1}^L \int _{-\infty}^{+\infty} \frac{dt}{2\pi }e^{itx_h}\hat \phi _H (k,t) + \nonumber \\
&+& \frac{ik}{2\pi |k|} e^{\frac{|k|}{2}} \int _{-\infty}^{+\infty} \frac{dt}{4\pi ^2} \hat \phi _H (k,t) [\hat \sigma (t) -2it \hat L(t) ]
 \, . \label {Seq1}
\ea
Then, we compute equation (\ref {Seq1}) at $k=0$. We obtain
\be S(0) =
-g^2 \int _{-\infty}^{+\infty} \, \frac{dt}{8\pi ^2 } \, \hat e(t) \, [ \hat \sigma (t)-2it\hat {L}(t) ] -  g^2 \sum _{h=1}^L \int _{-\infty}^{+\infty} \frac{dt}{4\pi} \, e^{itx_h} \, \hat e(t) \, , \label {k0}
\ee
where
\be
\hat e(t)=\frac{2 \sqrt{2}\, \pi }{gt} \, e^{-\frac{|t|}{2}} \, J_1 (\sqrt{2}gt) \label {et}
\ee
is the Fourier transform of
\be
e(u)=q_2(u)= i \left [ \frac {1}{x^+(u)} - \frac {1}{x^-(u)} \right ] \, . \label {ev}
\ee
On the other hand, if we want to compute the anomalous dimension $\gamma  (g,L,s)$ of a state described by a solution of the ABA equations, we have to compute the sum $\gamma = g^2 \sum \limits _{k=1}^s e (u_k)$.
Using formula (\ref {cauchy}) we obtain
\be
\gamma (g,L,s) = -g^2 \sum _{h=1}^{L} e(x_h)-g^2\int _{-\infty}^{+\infty} \frac {dv}{2\pi} e(v)
Z^{\prime}(v)+  g^2 \int _{-\infty}^{+\infty}\frac {dv}{\pi}
e(v)\frac {d}{dv} {\mbox {Im}} \ln
[1+(-1)^{L} \, e^{iZ(v-i0^+)}] \, ,
\ee
which can be written also, in terms of Fourier transforms, as
\be
\gamma (g,L,s)= - g^2 \int _{-\infty}^{+\infty} \ \frac{dt}{4\pi ^2 } \ \hat e(t) \ [ \hat \sigma (t)-2it\hat {L}(t) ] -
g^2 \sum _{h=1}^L \int _{-\infty}^{+\infty} \frac{dt}{2\pi} e^{itx_h} \hat e(t) \, . \label {gamma3}
\ee
Comparing (\ref {gamma3}) with (\ref {k0}), we gain relation (\ref {klip})

\medskip

For an alternative proof of (\ref {klip}), we first notice that:
\be
\hat \sigma (0) = \int_{-\infty}^{+\infty} du Z'(u)= -2\pi (L+s) \, . \label {sig0}
\ee
Then, using (\ref {Zlimit}) we find that, when $u \rightarrow \pm \infty$,
\be
L(u)\rightarrow  \frac{L+2s+\gamma (g,L,s)}{2u}+
O\left (\frac{1}{u^3}\right ) \, . \label {Llimits}
\ee
Using these property, one finds that
\be
\lim _{k\rightarrow 0^{\pm}} \hat L(k)=\mp \frac{i\pi}{2} [L+2s+\gamma (g,L,s)] \, , \label {Llimits2}
\ee
Inserting (\ref {sig0}, \ref {Llimits2}) into (\ref {Sdef}), we find again (\ref {klip}).
This alternative proof emphasizes the fact that the information on the anomalous dimension comes entirely from the term $\hat L(k)$,
which is a non-linear function of the counting function $Z(u)$:
\be
\lim _{k\rightarrow 0^{\pm}} \hat L(k)=\lim _{k\rightarrow 0^{\pm}} \int_{-\infty}^{+\infty} du e^{-iku} {\mbox {Im}} \ln
[1+(-1)^{L} \, e^{iZ(u-i0^+)}] = \pm \frac{\pi}{2i} [L+2s+\gamma (g,L,s)]\, . \label {gammafinal}
\ee
Curiously (and, perhaps, interestingly), formula (\ref {gammafinal}) looks similar
to the TBA expression for the free energy. We have checked formula (\ref {gammafinal}) in various particular cases
(twist $2,3$, one and two loops) for which explicit solutions \cite {BC} of ABA equations
was found \footnote {Results in this Appendix could seem in contrast with BES finding stating that anomalous dimension at order $\ln s$ is given by the value in zero of the Fourier transform of the 'higher than one loop density of roots', which satisfies BES equation. The solution to this apparent contrast is that the function $\hat \Sigma (k)$ which satisfies the BES equation and such that $\hat \Sigma (0^{+})=\pi \gamma (g,L,s) \Bigl |_{\ln s}$ in our notations reads
\be
\hat \Sigma (k)= \hat \sigma _H (k)  \Bigl |_{\ln s} + \frac{2 e^{-|k|}}{1-e^{-|k|}} \, ik \, \hat {L}_H(k) \Bigl |_{\ln s} \, , \label {sigma-BES}
\ee
where the label $H$ means that only higher than one loop contributions have to be included. Then, in addition to what we would call 'higher than one loop density of roots', {\it i.e.} $\hat \sigma _H (k)$, in BES density (\ref {sigma-BES}) there is also a term nonlinear in the counting function $Z$. This term can be estimated at large $s$ by using (\ref {nlt0}) to be (almost everywhere) $O(1/s^2)$: therefore, as far as the order $\ln s$ is concerned, it is almost everywhere negligible, with the exception of the point $k=0$, where one experiences the noncommutativity between the limits $s\rightarrow +\infty$ and $k\rightarrow 0$. However, the nonlinear term has to be kept in definition (\ref {sigma-BES}) of the BES density, since it gives the entire
information on anomalous dimensions (see (\ref {gammafinal})), due to the fact that $\hat \sigma _H (0)=\hat \sigma (0)-\hat \sigma |_{1\textrm{loop}}(0)=0$.}.

\section{Evaluation of $Z'(u)$ at high spin and large $u$} \label {Zprime}
\setcounter{equation}{0}

Using definition (\ref {Z}), we write the counting function $Z(u)$ in the region of large $u \sim b$:
\be
Z(u)= -L \pi + \frac{\gamma +L}{u} -2 \sum _{k=1}^s \arctan  (u-u_k) + O(1/b^2) \, . \nonumber
\ee
We add and subtract the sum over the internal holes and thus obtain
\be
Z(u)= -L \pi + \frac{\gamma +L}{u} -2 \sum _{k=1}^s \arctan  (u-u_k) - 2 \sum _{h=1}^{L-2} \arctan  (u-x_h)+ \pi (L-2) - \frac{2L-4}{u}+ O(1/b^2) \nonumber \, .
\ee
Then, the use of (\ref {cauchy}) gives
\be
Z(u)= -2\pi + \frac{\gamma -L+4}{u} + \int _{-b}^b \frac{dv}{\pi} \arctan (u-v) [Z'(v) -2L'(v)] + O(1/b^2) \, .
\ee
Evaluation of the nonlinear term is done using formula (2.17) of \cite {rep}:
\ba
Z(u) &=& -2\pi + \frac{\gamma -L+4}{u} + \int _{-b}^b \frac{dv}{\pi} \arctan (u-v) Z'(v) + \nonumber \\
&+& 2 \sum _{k=0}^{\infty} \frac{(2\pi)^{2k+1}}{(2k+2)!} B_{2k+2} (1/2) \left [ \frac{d}{dx^{2k+1}} \arctan (u-Z^{(-1)}(x) ) \right ]_{x=Z(-b)}^{x=Z(b)} +O(1/b^2)= \nonumber \\
&=& -2\pi + \frac{\gamma -L+4}{u} + \int _{-b}^b \frac{dv}{\pi} \arctan (u-v) Z'(v) + \frac{\pi}{6Z'(b)} \left [ \frac{1}{1+(u-b)^2}- \frac{1}{1+(u+b)^2} \right ] +O(1/b^2) \, , \nonumber
\ea
where we neglected higher order terms in the sum over $k$ since $Z'(b) \sim O(b)$ (\ref {Zprimeb}).
It is convenient to pass to the derivative of $Z$:
\be
Z'(u)= - \frac{\gamma -L+4}{u^2} + \int _{-b}^b \frac{dv}{\pi} \frac{1}{1+(u-v)^2} Z'(v)  +O(1/b^2) \label
{eqZprime} \, .
\ee
For large $u$, but still $0<u<b$, the solution to (\ref {eqZprime}) is
\be
Z'(u)=-2 \ln \frac{b+\sqrt{b^2-u^2}}{b-\sqrt{b^2-u^2}} +\pi (\gamma -L+4) \delta (u) + O(1/b^2) \label {Zprimein} \, .
\ee
Indeed, if we insert (\ref {Zprimein}) into the integral of (\ref {eqZprime}) and use
the integration formula
\be
\int _{-b}^{b} \frac{dv}{z-v} \ln \frac{b+\sqrt{b^2-v^2}}{b-\sqrt{b^2-v^2}}=
i\pi \textrm{sgn} ( \textrm{Im}  z) \ln \frac{\sqrt{b^2 -z^2}-b}{\sqrt{b^2 -z^2}+ b} \, , \quad z \not \in [-b,b] \, , \label {intZprime}
\ee
in the rhs of this last equation (with $z=u\pm i$), when $0<u<b$ we are left with
\be
- \ln \frac{\sqrt{b^2-(u+i)^2}+b}{\sqrt{b^2-(u+i)^2}-b} - \ln \frac{\sqrt{b^2-(u-i)^2}+b}{\sqrt{b^2-(u-i)^2}-b}+O(1/b^2)=  -2 \ln \frac{b+\sqrt{b^2-u^2}}{b-\sqrt{b^2-u^2}}  + O(1/b^2) \, ,
\ee
which matches (\ref {Zprimein}).
Plugging approximation (\ref {Zprimein}) into equation (\ref {eqZprime}) and letting $u>b$, we find $Z'(u)$ in this domain. Application of (\ref {intZprime}) gives
\be
u>b \, \quad \, , \quad Z'(u)= - \ln \frac {\sqrt{b^2-(u+i)^2}+b}{\sqrt{b^2-(u+i)^2}-b} - \ln \frac {\sqrt{b^2-(u-i)^2}+b}{\sqrt{b^2-(u-i)^2}-b} +O(1/b^3) \, ,
\ee
which, since $u >b \gg 1$, is expanded as follows
\be
Z'(u)=- \frac{4b}{u} \frac{1}{\sqrt{u^2-b^2}} + O(1/b^3) \, . \label {outZprime}
\ee

\section{Neumann expansion for $S(k)$} \label{neu_exp}
\setcounter{equation}{0}
Let see how the Neumann expansion for $S(k)$ works. This is a standard procedure \cite {KL} in the case of an integral equation with separable kernel:
\ba
&& S(k)=\sum _{p=1}^{\infty} \mathcal{S}_p (g,L,s) \frac {J_p({\sqrt 2}gk)}{k} \, , \quad
\mathcal{S}_p (g,L,s)=\sum _{n=-1}^{\infty} \left [ S_p^{(n)}(g,L)+\frac{1}{s} \check S_p^{(n)}(g,L) + O(1/s^2) \right ] (\ln s )^{-n}
\Rightarrow \nonumber \\
&& \gamma^{(n)}(g,L) = {\sqrt 2} g S_1^{(n)}(g,L) \, , \ \ \check \gamma^{(n)}(g,L) = {\sqrt 2} g \check S_1^{(n)}(g,L) \, .
\ea
The Neumann expansion transforms the linear integral equation for $S(k)$ into a set of linear infinite system. In particular,
$\check S_p^{(n)}(g,L)$, $n \geq 1$, satisfy the system\footnote {We use the notation ($J_n$ is a Bessel function):
\begin{equation}
Z_{n,m}(g)= \int _{0}^{+\infty}
\frac {dt}{t} \frac {J_{n}({\sqrt {2}}gt)J_{m}({\sqrt {2}}gt)}{e^t-1}
\, .
\end{equation}},
\begin{eqnarray}
\check S^{(n)}_{2p-1}(g,L)&=& \sqrt{2} g \delta _{p,1} \gamma ^{(n)}(g,L) -(2p-1) \int _{0}^{+\infty}
\frac{dt}{t} \frac {\check {P}^{(n)}(g,t) \, J_{2p-1}({\sqrt {2}}gt)}{\sinh \frac {t}{2}} - \nonumber \\
&-& 2(2p-1)
\sum _{m=1}^{\infty} Z_{2p-1,m}(g) \check S^{(n)}_m (g,L) \, , \label {Ssystem} \\
\check S^{(n)}_{2p}(g,L)&=& - 2p \int _{0}^{+\infty}
\frac{dt}{t} \frac { \check {P}^{(n)}(g,t) \,   J_{2p}({\sqrt {2}}gt)}{\sinh \frac {t}{2}} -
4p\sum _{m=1}^{\infty} Z_{2p,m}(g) (-1)^m \check S^{(n)}_{m}(g,L) \, , \nonumber
\end{eqnarray}
where $\check {P}^{(n)}(g,t)$ appears in the expansion
\be
P(s,g,t)=\sum _{h=1}^{L-2} \left[ \cos t x_h -1 \right]=   \sum_{n=1}^\infty \left ( P^{(n)} (g,t)
+ \frac{\check {P}^{(n)}(g,t)}{s}  \right )(\ln s )^{-n} + O\left ( s^{-1}(\ln s)^{-\infty} \right ) \, ,
\label {logexpan}
\ee
which follows from (\ref {upos}). More explicitly,
inserting (\ref {upos}) into (\ref {logexpan}) we obtain
\ba
\check {P}^{(n)} (g,t) &=&
\sum_{r=1}^n  t^r \; \cos \frac{\pi r}{2}\sum_{\{ j_1, \dots, j_{n-r+1}\}}
\frac{ \sum\limits_{h=1}^{L-2}\left (\prod\limits_{m=1}^{n-r+1} (\alpha_{m,h})^{j_m}\right )\sum\limits_{m'=1}^{n-r+1} j_{m'}\frac{\check \alpha _{m',h}}{\alpha _{m',h}} }{\prod\limits_{m=1}^{n-r+1}  j_m !}\, , \nonumber \\
&& \sum _{m=1}^{n-r+1}j_m=r \, , \ \ \ \sum _{m=1}^{n-r+1}m j_m=n \, . \label {checkP}
\ea
Plugging (\ref {checkP}) into (\ref {Ssystem}) we obtain ($n \geq 1$)
\ba
&&\check S^{(n)}_p(g,L) = -2 \pi \sum_{r=1}^n  \tilde S^{(r/2)}_{p}(g) \; \cos \frac{\pi r}{2}  \sum_{\{ j_1, \dots, j_{n-r+1}\}}
\frac{ \sum\limits_{h=1}^{L-2}\left (\prod\limits _{m=1}^{n-r+1} (\alpha_{m,h})^{j_m}\right ) \sum\limits _{m'=1}^{n-r+1} j_{m'}\frac{\check \alpha _{m',h}}{\alpha _{m',h}}}{\prod\limits_{m=1}^{n-r+1}  j_m !}  + \nonumber \\
&+& \frac{S^{(-1)}_p(g)}{2}
 \gamma^{(n)} (g,L) \, \, , \, \quad \sum _{m=1}^{n-r+1}j_m=r \, , \,   \sum _{m=1}^{n-r+1}m j_m=n \, , \label {gammamain}
\ea
where the 'reduced coefficients' $\tilde S^{(r)}_{p}(g)$ satisfy the systems (see \cite {FRS6})
\begin{eqnarray}
\tilde S^{(r)}_{2p-1}(g)&=& {\mathbb I}_{2p-1}^{(r)}(g) - 2(2p-1)
\sum _{m=1}^{\infty} Z_{2p-1,m}(g) \tilde S^{(r)}_m (g) \, , \nonumber \\
\tilde S^{(r)}_{2p}(g)&=& {\mathbb I}_{2p}^{(r)}(g) -
4p\sum _{m=1}^{\infty} Z_{2p,m}(g) (-1)^m \tilde S^{(r)}_{m}(g) \, , \label {tildeSsystem}
\end{eqnarray}
with
\begin{equation}
\label{intforterm}
{\mathbb I}_p^{(r)}(g)= p \int _{0}^{+\infty} \frac {dh}{2\pi} h^{2r-1}
\frac {J_p ({\sqrt {2}}gh)}{\sinh \frac {h}{2}} \, .
\end{equation}

\section{Solution of internal holes equation: $\check \alpha_{n,h}$ expressed in terms of the densities and their derivatives} \label{alpha_den}
\setcounter{equation}{0}
We want to extract from (\ref {condhole}), explicit expressions for $\alpha_{m,h}$ and $\check \alpha_{m,h}$ in terms of the densities and their derivatives, using the notation (\ref{defsigmarn}). We can use techniques and results of \cite {FGR5}. In particular, relations (2.8) of
\cite {FGR5} are still valid, after substituting in them
\be
\alpha_{m,h} \rightarrow  \alpha_{m,h} + \frac{\check \alpha_{m,h}}{s} \, , \quad \sigma ^{(n)}_r\rightarrow
\sigma ^{(n)}_r+\frac{\check \sigma ^{(n)}_r}{s} \, .
\ee
We can then solve for $\check \alpha_{m,h}$. We obtain
\begin{eqnarray}
\check \alpha_{p+1,h}  &  =  & -  \sum_{r=1}^{p}\left[ \frac{\sigma^{(-1)}_{r}}{ \sigma^{(-1)}_0}
\sum_{m'=1}^{p-r+1} \frac{j_{m'}(\check \alpha_{m',h})}{(\alpha_{m',h})}+\frac{\check \sigma^{(-1)}_{r}}{ \sigma^{(-1)}_0}
-\frac{\sigma^{(-1)}_{r}\check \sigma^{(-1)}_0}{( \sigma^{(-1)}_0)^2}\right]\sum_{\{ j_1, \dots, j_{p-r+1}\}}
\prod_{m=1}^{p-r+1} \frac{(\alpha_{m,h})^{j_m}}{j_m !}+ \nonumber \\
&&-\sum_{l=0}^{p-1}   \sum_{r=1}^{p-l}\left[ \frac{\sigma^{(l)}_{r-1}}{ \sigma^{(-1)}_0}
\sum_{m'=1}^{p-r-l+1} \frac{j_{m'}(\check \alpha_{m',h})}{(\alpha_{m',h})}+\frac{\check \sigma^{(l)}_{r-1}}{ \sigma^{(-1)}_0}
-\frac{\sigma^{(l)}_{r-1}\check \sigma^{(-1)}_0}{( \sigma^{(-1)}_0)^2}\right]\sum_{\{ j_1, \dots, j_{p-r-l+1}\}}
\prod_{m=1}^{p-r-l+1} \frac{(\alpha_{m,h})^{j_m}}{j_m !} \nonumber \\
\check \alpha_{1,h}  &  =  & \frac{-\pi\check \sigma_0^{(-1)}(2h+1-L) }{(\sigma^{(-1)}_0)^2} \, ,
\end{eqnarray}
where the coefficients $j_m$ of the first term in the right hand side satisfy $ \sum\limits_{m=1}^{p-r+1} j_m=r+1$, $\sum\limits_{m=1}^{p-r+1} mj_m=p+1$, while the coefficients $j_m$ related to the second term satisfy $\sum\limits_{m=1}^{p-r-l+1} j_m=r$,
$\sum\limits_{m=1}^{p-r-l+1} mj_m=p-l$.
The first $\check \alpha_{m,h}$ are
\ba
&& \check \alpha_{1,h}= -\frac{\pi (2h+1-L) \check \sigma ^{(-1)}_0}{ (\sigma^{(-1)}_0)^2 } \, , \quad
\check \alpha_{2,h}= -\frac{\pi (2h+1-L)\check \sigma^{(0)}_0}{  (\sigma^{(-1)}_0)^2 }
+\frac{2\pi (2h+1-L) \sigma ^{(0)}_0 \check \sigma ^{(-1)}_0}{(\sigma ^{(-1)}_0)^3} \, , \nonumber \\
&& \check \alpha_{3,h}= \pi (2h+1-L) \left [ \frac{2 \sigma ^{(0)}_0 \check \sigma ^{(0)}_0}{(\sigma ^{(-1)}_0)^3}
- \frac{3 (\sigma ^{(0)}_0)^2 \check \sigma ^{(-1)}_0}{(\sigma ^{(-1)}_0)^4} \right ]- \frac{\pi ^3 (2h+1-L)^3}{6} \left [ \frac{\check {\sigma}^{(-1)}_2}{(\sigma ^{(-1)}_0)^4} -\frac{4 {\sigma}^{(-1)}_0\check \sigma ^{(-1)}_0}{(\sigma ^{(-1)}_0)^5} \right ]\, , \nonumber\\
&& \check \alpha_{4,h}= \pi (2h+1-L) \left [\frac{4 (\sigma ^{(0)}_0)^3 \check \sigma ^{(-1)}_0}{(\sigma ^{(-1)}_0)^5} -\frac{3 (\sigma ^{(0)}_0)^2 \check \sigma ^{(0)}_0}{(\sigma ^{(-1)}_0)^4} +\frac{2 \sigma ^{(2)}_0 \check \sigma ^{(-1)}_0}{(\sigma ^{(-1)}_0)^3} -\frac{\check \sigma ^{(2)}_0}{(\sigma ^{(-1)}_0)^2}\right]+ \label {alfa}\\
&&+\frac{2\pi ^3 (2h+1-L)^3}{3}\left[-\frac{5\sigma_2^{(-1)}\sigma_0^{(0)}\check \sigma_0^{(-1)}}{(\sigma_0^{(-1)})^6}+\frac{\sigma_0^{(0)}\check \sigma_2^{(-1)}}{(\sigma_0^{(-1)})^5} +\frac{\sigma_2^{(-1)}\check \sigma_0^{(0)}}{(\sigma_0^{(-1)})^5}+\frac{\sigma_2^{(0)}\check \sigma_0^{(-1)}}{(\sigma_0^{(-1)})^5}-\frac{\check \sigma_2^{(0)}}{4(\sigma_0^{(-1)})^4}\right] \nonumber
\, .
\ea
Now, inserting (\ref{alfa}) into (\ref{gamma 234}) we can derive expressions for $\check S^{(n)} _p(g,L)$, in terms of $\sigma^{(n)}_r$ and  $\check{\sigma}^{(n)}_r$: for $n=2,...,5$ we obtain,
\begin{eqnarray}
\check S^{(2)}_p (g,L)&=&\frac{S_p^{(-1)}(g)}{2}\gamma^{(2)} (g,L)-\frac{2\pi^3}{3}\tilde{S}_p^{(1)} (g)\frac{\check \sigma^{(-1)}_0}{\left(\sigma^{(-1)}_0\right)^3}(L-1)(L-2)(L-3) \, , \nonumber\\
\check S^{(3)}_p(g,L) &=&\frac{S_p^{(-1)}(g)}{2}\gamma^{(3)} (g,L)+\frac{2\pi^3}{3}\frac{\tilde{S}_p^{(1)} (g)}{\left(\sigma^{(-1)}_0\right)^3}\left[\frac{3\sigma^{(0)}_0\check \sigma^{(-1)}_0}{\sigma^{(-1)}_0}-\check \sigma^{(0)}_0\right](L-1)(L-2)(L-3)\nonumber \, , \\
\check S^{(4)}_p(g,L) &=&\frac{S_p^{(-1)}(g)}{2}\gamma^{(4)} (g,L)+\frac{2\pi^3}{3}\frac{1}{\left(\sigma^{(-1)}_0\right)^4}\left\{\frac{\pi^2}{6\sigma^{(-1)}_0}\left[\tilde{S}^{(1)}_p (g)\left(\frac{5\sigma^{(-1)}_2 \check \sigma^{(-1)}_0}{\sigma^{(-1)}_0}-\check \sigma^{(-1)}_2\right)+\right.\right.\nonumber\\
&&\left.\left. +\check \sigma^{(-1)}_0 \tilde{S}^{(2)}_p (g)\frac{}{}\right]\frac{(5+3L(L-4))}{5}+3\sigma^{(0)}_0\tilde{S}^{(1)}_p (g)\left[\check \sigma_0^{(0)}-\frac{2\sigma^{(0)}_0 \check \sigma^{(-1)}_0}{\sigma^{(-1)}_0}\right]\right\}\times\nonumber\\
&&\times(L-1)(L-2)(L-3)\label{gamma 234 sigma} \, , \\
\check S^{(5)}_p (g,L) &=&\frac{S_p^{(-1)}(g)}{2}\gamma^{(5)} (g,L)+\frac{2\pi^3}{3}\frac{1}{\left(\sigma^{(-1)}_0\right)^3}\left\{\frac{\pi^2}{2\left(\sigma^{(-1)}_0\right)^2}\left[\tilde{S}^{(1)}_p (g)\left(\frac{\sigma^{(0)}_2 \check \sigma^{(-1)}_0}{\sigma^{(-1)}_0} +\frac{\sigma^{(0)}_0 \check \sigma^{(-1)}_2}{\sigma^{(-1)}_0}+\frac{\sigma^{(-1)}_2 \check \sigma^{(0)}_0}{\sigma^{(-1)}_0}-\right.\right.\right.\nonumber\\
&&\left. \left. - \frac{\check \sigma^{(-1)}_2 }{5}-\frac{6\sigma_2^{(-1)}\sigma_0^{(0)}\check \sigma_0^{(-1)}}{\left(\sigma_0^{(-1)}\right)^2}\right)
+\tilde{S}^{(2)}_p (g)\left(5\check \sigma_0^{(0)} -\frac{\sigma_0^{(0)}\check \sigma_0^{(-1)}}{\sigma_0^{(-1)}}\right)\right]\frac{(5+3L(L-4))}{3}+\nonumber\\
&&\left. \tilde{S}^{(1)}_p (g)\left[-\check \sigma_0^{(2)}+\frac{3\sigma^{(2)}_0 \check \sigma^{(-1)}_0}{\sigma^{(-1)}_0} -\frac{6(\sigma^{(0)}_0)^2 \check \sigma^{(0)}_0}{(\sigma^{(-1)}_0)^2}+\frac{10(\sigma^{(0)}_0)^3 \check \sigma^{(-1)}_0}{(\sigma^{(-1)}_0)^3}\right]\right\}\times\nonumber\\
&&\times(L-1)(L-2)(L-3)\nonumber \, .
\end{eqnarray}

\section{Useful relations} \label {sigf}
\setcounter{equation}{0}

It is possible to express certain ratios among coefficients $\sigma_r^{(n)}$ and $\check \sigma_r^{(n)}$ in terms of functions $f(g)$, $f_{sl} (g,L)$ and $\gamma^{(n)} (g,L)$.
We now find some of these relations, which are useful in order to prove (\ref {Srec1}-\ref {Srec5}).

Let us start with equation (2.14) of \cite{FGR5} and equation (\ref{checkP}). Comparing them, we find the following relation:
\be \label{relP}
\check{{P}}^{(2)} (g,t)=-2\frac{\check \sigma_0^{(-1)}}{\sigma_0^{(-1)}}{P}^{(2)} (g,t).
\ee
Now, developing $S(k)$ according to relation (\ref {Shigh}),
\be \label{Sdev}
S(k)=\sum_{n=-1}^{\infty} S^{(n)} (k) {(\ln s)}^{-n} + \sum_{n=-1}^{\infty} \check S^{(n)} (k) \frac{{(\ln s)}^{-n}}{s}  + O\left ( s^{-1}(\ln s)^{-\infty} \right ) \, ,
\ee
and using integral equation (\ref{Sdef2}) together with equation (\ref{gamma}), it is possible to obtain the following relations:
\begin{eqnarray}
\check S^{(-1)} (k) &=& \frac{f(g)}{2} S^{(-1)} (k)\label{relS1} \, , \\
\check S^{(0)} (k) &=& \frac{f_{sl} (g,L)+ L-1}{2}  S^{(-1)} (k)\label{relS2} \, , \\
\check S^{(2)} (k) &=& \frac{\gamma^{(2)} (g,L)}{2} S^{(-1)} (k) -f(g) S^{(2)} (k) \, . \label{relS3}
\end{eqnarray}
For what concerns $\hat{\sigma} (k)$ we have the exact expression:
\begin{eqnarray} \label{sigmak2}
\hat{\sigma} (k)&=& - \frac{2\pi L e ^{-\frac{|k|}{2}}}{1-e^{-|k|}}+ \frac{2\pi L e ^{-|k|}}{1-e^{-|k|}}
+ \frac{2\pi e ^{-|k|}}{1-e^{-|k|}} \sum _{h=1}^L \left ( \cos kx_h -1 \right )- \frac{2 ik e ^{-|k|}}{1-e^{-|k|}} \hat L(k) + \hat G(k)  \, ,
\end{eqnarray}
where
\be
\hat G(k)=\frac{\pi |k|}{\sinh \frac{|k|}{2}} S(k) \, .
\ee
Then, applying inverse Fourier transform, we obtain:
\ba
\sigma (u)&=& L \left [ \psi \left ( \frac{1}{2}-iu \right ) + \psi \left ( \frac{1}{2}+iu \right ) \right ]
- (L-2) [ \psi (1-iu) +\psi (1+iu)] - \nonumber \\
&-& \psi (1-ix_L-iu) - \psi (1+ix_L-iu) - \psi (1-ix_L+iu) - \psi (1+ix_L+iu) -
\left [ 2\ln 2 +  O \left ( \frac{u^2}{s^2} \right )  \right ] + \nonumber \\
&+& \int _{-\infty}^{+\infty} dk e^{iku} \frac{e^{-|k|}}{1-e^{-|k|}} P(s,g,k) + G(u) \, . \label {sigapp}
\ea
Using the position of the external holes (\ref {extholes}) and computing $\sigma (u)$ at $u=0$, we obtain
\begin{eqnarray}
\sigma (0)&=& -4 \ln s -4\gamma_E -4L \ln 2 +G(0) -2f(g)\frac{\ln s}{s}-2\frac{f_{sl} (g,L)+L-1}{s}+\label{sigmauzero} \\
&&-2\sum_{n=1}^{\infty}\frac{\gamma^{(n)} (g,L)}{s} (\ln s)^{-n}+\sum_{n=1}^{\infty}\int _{-\infty}^{+\infty} dk \frac{e^{-|k|}}{1-e^{-|k|}}\left( P^{(n)} (g,k) + \frac{\check{P}^{(n)} (g,k)}{s}\right){(\ln s)}^{-n}  + O\left ( s^{-1}(\ln s)^{-\infty} \right ) . \nonumber
\end{eqnarray}
It is obvious that, expanding $G(0)$ in the same way of $S(k)$ in (\ref{Sdev}), relations (\ref{relS1}), (\ref{relS2}) and (\ref{relS3}) are
also valid for the corresponding coefficients of $G(0)$.
Using these relations and also (\ref{relP}) it is possible to find, from (\ref{sigmauzero}) and remembering (\ref{defsigmarn}), the following relations:
\be \label{apprelfg}
f(g)=2\frac{\check \sigma_0^{(-1)}}{\sigma_0^{(-1)}} \, ,
\quad
f_{sl} (g,L)=2 \frac{\check \sigma_0^{(0)}}{\sigma_0^{(-1)}} - (L-1) \, ,
\quad
\gamma^{(2)} (g,L)= 2\frac{\check \sigma_0^{(2)}}{\sigma_0^{(-1)}}+4\frac{\sigma_0^{(2)} \check \sigma_0^{(-1)}}{{(\sigma_0^{(-1)})}^2} \, .
\ee
Computing from (\ref {sigapp}) the second derivative of $\sigma (u)$ at $u=0$, it is also possible to show that:
\be \label{apprelfg2}
f(g)=2\frac{\check \sigma_2^{(-1)}}{\sigma_2^{(-1)}} \, .
\ee

\section{Explicit expressions for $S^{(n)}_p (g,L)$ with $n=1,2,3,4,5$} \label {gamman}
\setcounter{equation}{0}

We report here the expressions of the functions $S^{(n)}_p (g,L)$ with $n=1,2,3,4,5$ in terms of the densities (and their derivatives in zero) and the solutions of the 'reduced systems' $\tilde S_p^{(n)}(g)$. The general method to obtain them and results for $n=1,2,3,4$ are shown in \cite{FGR5}.
\ba
S^{(1)}_p (g,L) &=&0 \, , \\
S^{(2)}_p (g,L) &=& \frac{\pi^3}{3(\sigma_0^{(-1)})^2} (L-3)(L-2)(L-1)\tilde S_p^{(1)} (g) \, , \\
S^{(3)}_p (g,L) &=& -2 \frac{\pi^3 \sigma_0^{(0)}}{3(\sigma_0^{(-1)})^3} (L-3)(L-2)(L-1)\tilde S_p^{(1)} (g) \, , \\
S^{(4)}_p (g,L) &=& 2\pi (L-3)(L-2)(L-1)\left\{ \left[\frac{\pi^2{\left(\sigma_0^{(0)}\right)}^2}{2(\sigma_0^{(-1)})^4}-\frac{\pi^4\sigma_2^{(-1)}}{90(\sigma_0^{(-1)})^5}(5+3L(L-4))\right]\tilde S_p^{(1)} (g)\right.\nonumber\\
&& \left. -\frac{\pi^4}{360(\sigma_0^{(-1)})^4}(5+3L(L-4))\tilde S_p^{(2)} (g)\right\}\, , \\
S^{(5)}_p (g,L) &=& (L-3)(L-2)(L-1)\left\{\left[\left(\frac{5\pi^5}{3}\frac{\sigma_2^{(-1)}\sigma_0^{(0)}}{{\left(\sigma_0^{(-1)}\right)}^6}-\frac{\pi^5}{3}\frac{\sigma_2^{(0)}}{{\left(\sigma_0^{(-1)}\right)}^5}\right)\frac{(5+3L(L-4))}{15}+\right.\right.\\
&&+\left.\left. \left(-\frac{4\pi^3{\left(\sigma_0^{(0)}\right)}^3}{3{\left(\sigma_0^{(-1)}\right)}^5}-\frac{2\pi^3\sigma_0^{(2)}}{3{\left(\sigma_0^{(-1)}\right)}^3}\right)\right]\tilde S_p^{(1)} (g)+\frac{\pi^5\sigma_0^{(0)}}{45{\left(\sigma_0^{(-1)}\right)}^5}\tilde S_p^{(2)} (g) (5+3L(L-4))\right\} \, . \nonumber
\ea



\begin{thebibliography}{xx}

\bibitem{MGKPW1}
J.M. Maldacena, {\sl The large N limit of superconformal field
theories and supergravity}, Adv. Theor. Math. Phys.
{\bf 2} (1998) 231 and hep-th/9711200;
\bibitem{MGKPW2}
S.S. Gubser, I.R. Klebanov, A.M. Polyakov, {\sl Gauge theory
correlators from non-critical string theory},
Phys.Lett. {\bf B428} (1998) 105 and hep-th/9802109;
\bibitem{MGKPW3}
E. Witten, {\sl Anti-de Sitter space and holography}, Adv. Theor.
Math. Phys. {\bf 2} (1998) 253 and hep-th/9802150;

\bibitem{MZ}
J. Minahan, K. Zarembo,
{\sl The Bethe Ansatz for ${\cal N}=4$
Super Yang-Mills}, JHEP{\bf 03} (2003) 013 and hep-th/0212208;

\bibitem{BS1}
N. Beisert, M. Staudacher,
{\sl The ${\cal N}=4$ SYM integrable super spin
chain}, Nucl. Phys. {\bf B670} (2003) 439 and hep-th/0307042;
\bibitem{BS2}
V. Kazakov, A. Marshakov, J. Minahan, K. Zarembo,
{\sl Classical/quantum integrability in AdS/CFT},
JHEP{\bf 05} (2004) 024 and hep-th/0402207;
\bibitem{BS3}
M. Staudacher, {\sl The factorized S-matrix of CFT/AdS},
JHEP{\bf 05} (2005) 054 and hep-th/0412188;
\bibitem{BS4}
N. Beisert, V. Kazakov, K. Sakai, K. Zarembo,
{\sl The Algebraic curve of classical superstrings on AdS(5) x S**5},
Commun. Math. Phys. {\bf 263} (2006) 659 and hep-th/0502226;
\bibitem{BS5}
N. Beisert, M. Staudacher,
{\sl Long-range $PSU(2,2|4)$ Bethe Ansatz
for gauge theory and strings},
Nucl. Phys. {\bf B727} (2005) 1 and
hep-th/0504190;

\bibitem{BES}
N. Beisert, B.Eden, M. Staudacher, {\sl Transcendentality and
crossing}, J.Stat.Mech.{\bf 07} (2007) P01021 and hep-th/0610251;

\bibitem{WRA1}
J. Ambjorn, R. Janik, C. Kristjansen,
{\sl Wrapping interactions and a new source of
corrections to the spin chain/string duality},
Nucl. Phys. {\bf B736} (2006) 288
and hep-th/0510171;
\bibitem{WRA2}
A. Kotikov, L. Lipatov, A. Rej, M. Staudacher, V. Velizhanin,
{\sl Dressing and wrapping},
J. Stat. Mech. {\bf 10} (2007) P003 and
arXiv:0704.3586 [hep-th];

\bibitem{TBA1}
D. Bombardelli, D. Fioravanti and R. Tateo,
{\sl Thermodynamic Bethe Ansatz for planar AdS/CFT: a proposal},
 J. Phys. A  {\bf 42} (2009) 375401
and arXiv:0902.3930 [hep-th];
\bibitem{TBA2}
N. Gromov, V. Kazakov, A. Kozak and P. Vieira,
{\sl Exact Spectrum of Anomalous Dimensions of Planar N = 4 Supersymmetric Yang-Mills Theory: TBA and excited states},
Lett. Math. Phys. {\bf 91} (2010) 265, cf. also
arXiv:0902.4458 [hep-th];
\bibitem{TBA3}
G. Arutyunov and S. Frolov,
{\sl Thermodynamic Bethe Ansatz for the $AdS_5 \times  S^5$ Mirror Model},
JHEP{\bf 05} (2009) 068
and arXiv:0903.0141 [hep-th];
\bibitem{TBA4}
D. Bombardelli, D. Fioravanti and R. Tateo,
{\sl TBA and Y-system for planar $AdS_4/CFT_3$},
Nucl. Phys. {\bf B834} (2010) 543 and arXiv:0912.4715 [hep-th];
\bibitem{TBA5}
N. Gromov and F. Levkovich-Maslyuk,
{\sl Y-system, TBA and Quasi-Classical strings in AdS(4) x CP3},
JHEP{\bf 06} (2010) 88 and
arXiv:0912.4911 [hep-th];

\bibitem{Y}
N. Gromov, V. Kazakov, P. Vieira,
{\sl Exact Spectrum of Anomalous Dimensions of Planar N=4 Supersymmetric Yang-Mills Theory},
Phys. Rev. Lett. {\bf 103} (2009) 131601 and arXiv:0901.3753 [hep-th];

\bibitem{CFT}
A. Cavagli\`{a}, D. Fioravanti, R. Tateo,
{\sl Extended Y-system for the $AdS_5/CFT_4$ correspondence},
Nucl. Phys. {\bf B843} (2011) 302 and arXiv:1005.3016 [hep-th];

\bibitem{GKPII}
S. Gubser, I. Klebanov and A. Polyakov,
{\sl A semiclassical limit of the gauge/string correspondence},
Nucl.Phys. {\bf B636} (2002) 99 and hep-th/0204051;

\bibitem{FT}
S. Frolov and A. Tseytlin,
{\sl Semiclassical quantization of rotating superstring in AdS(5) x S(5)},
  JHEP {\bf 0206} (2002) 007 and hep-th/0204226;


\bibitem{FPR1}
D. Fioravanti, S. Piscaglia, M. Rossi,
{\sl On the scattering over the GKP vacuum},
Phys. Lett. {\bf B728} (2014) 288-295 and arXiv:1306.2292 [hep-th];
\bibitem{FPR2}
D. Fioravanti, S. Piscaglia, M. Rossi,
{\sl Asymptotic Bethe Ansatz on the GKP vacuum as a defect spin chain: scattering, particles and minimal area Wilson loops}, Nucl. Phys. {\bf B898} (2015) 301 and arXiv:1503.08795 [hep-th];


\bibitem{BFKL}
L.N. Lipatov,
{\sl Reggeization of the Vector Meson and the Vacuum Singularity in Nonabelian Gauge Theories},
Sov. J. Nucl. Physics {\bf 23} (1976) 338-345 and Yad. Fiz. {\bf 23} (1976) 642-656;

\bibitem{MVV1}
S. Moch, J. A. M. Vermaseren and A. Vogt,
{\sl The Three loop splitting functions in QCD: The Nonsinglet case},
Nucl. Phys. {\bf B688} (2004) 101 and hep-ph/0403192;
\bibitem{MVV2}
A. Vogt, S. Moch and J. A. M. Vermaseren,
{\sl The Three-loop splitting functions in QCD: The Singlet case},
Nucl. Phys. {\bf B691} (2004) 129
and hep-ph/0404111;

\bibitem{DMS1}
Yu.L. Dokshitzer, G. Marchesini, G.P. Salam,
{\sl Revisiting parton evolution and the large-x limit},
Phys.Lett. {\bf B634} (2006) 504 and hep-ph/0511302;
\bibitem{DMS2}
B. Basso, G.P. Korchemsky, {\sl Anomalous dimensions of high-spin operators beyond the leading order},
Nucl. Phys. {\bf B775} (2007) 1 and hep-th/0612247;



\bibitem{REC1}
Yu.L. Dokshitzer, G. Marchesini, G.P. Salam,
{\sl N=4 SUSY Yang-Mills: three loops made simple(r)},
Phys.Lett. {\bf B646} (2007) 189 and hep-th/0612248;
\bibitem{REC2}
M. Beccaria, Yu.L. Dokshitzer, G. Marchesini,
{\sl Twist 3 of the sl(2) sector of N=4 SYM and reciprocity respecting evolution},
Phys.Lett. {\bf B652} (2007) 194-202 and arXiv:0705.2639 [hep-th];
\bibitem{REC3}
M. Beccaria, V.Forini,
{\sl Reciprocity of gauge operators in N=4 SYM},
JHEP{\bf 06} (2008) 077 and arXiv:0803.3768 [hep-th];
\bibitem{REC4}
V. Forini, M. Beccaria,
{\sl QCD-like properties for anomalous dimensions in N=4 SYM},
Theor. Math. Phys. {\bf 159} (2009) 712 and arXiv:0810.0101 [hep-th];
\bibitem{REC5}
M. Beccaria, V.Forini,
{\sl Four loop reciprocity of twist two operators in N=4 SYM},
JHEP{\bf 03} (2009) 111 and arXiv:0901.1256 [hep-th];
\bibitem{REC6}
M. Beccaria, G. Macorini,
{\sl Reciprocity and integrability in the sl(2) sector of N=4 SYM},
JHEP{\bf 01} (2010) 031 and arXiv:0910.4630 [hep-th];
\bibitem{REC7}
M. Beccaria, V.Forini, G. Macorini,
{\sl Generalized Gribov-Lipatov Reciprocity and AdS/CFT},
Adv. High Energy Phys. (2010) 753248 and arXiv:1002.2363 [hep-th];
\bibitem{REC8}
G. Georgiou, G. Savvidy,
{\sl Large spin behavior of anomalous dimensions and short-long strings duality},
J.Phys. {\bf A44} (2011) 305402 and arXiv:1012.5580 [hep-th];
\bibitem{REC9}
M. Beccaria, G. Macorini, C.A. Ratti
{\sl Wrapping corrections, reciprocity and BFKL beyond the sl(2) subsector in N=4 SYM},
JHEP{\bf 06} (2011) 071 and arXiv:1105.3577 [hep-th];

\bibitem{ABL}
L. Alday, A. Bissi, T. Lukowski,
{\sl Large spin systematics in CFT},
arXiv:1502.07707 [hep-th];

\bibitem{BFR}
D. Bombardelli, D. Fioravanti, M. Rossi, {\sl Large spin
corrections in ${\cal N}=4$ SYM $sl(2)$: still a linear integral
equation}, Nucl. Phys. {\bf B810} (2009) 460 and arXiv:0802.0027 [hep-th];

\bibitem{BJL1}
Z. Bajnok, R. Janik, T. Lukowski,
{\sl Four loops twist two, BFKL, wrapping and strings},
Nucl. Phys. {\bf B816} (2009) 376 and
arXiv:0811.4448 [hep-th];
\bibitem{BJL2}
T. Lukowski, A. Rej, V. Velizhanin,
{\sl Five-loops anomalous dimension of twist-two operators},
Nucl. Phys. {\bf B831} (2010) 105 and
arXiv:0912.1624 [hep-th];
\bibitem{BJL3}
V. Velizhanin,
{\sl Six-Loop Anomalous Dimension of Twist-Three Operators in N=4 SYM},
JHEP{\bf 11} (2010) 129 and arXiv:1003.4717 [hep-th];

\bibitem{AFS1}
G. Arutyunov, S. Frolov, M. Staudacher,
{\sl  Bethe ansatz for quantum strings},
JHEP{\bf 10} (2004) 016 and hep-th/0406256;
\bibitem{AFS2}
N. Beisert, R. Hernandez, E. Lopez,
{\sl A Crossing-symmetric phase for AdS(5) x S**5 strings},
JHEP{\bf 11} (2006) 070 and hep-th/0609044;

\bibitem{NLIE}
C. Destri, H.J. de Vega,
{\sl Unified approach to thermodynamic Bethe Ansatz and finite size corrections for lattice models and field theories},
Nucl. Phys. {\bf  B438} (1995) 413 and
hep-th/9407117;

\bibitem{BGK}
A. Belitsky, A. Gorsky, G. Korchemsky,
{\sl Logarithmic scaling in gauge/string correspondence},
Nucl. Phys. {\bf B748} (2006) 24 and hep-th/0601112;

\bibitem{FRS1}
L. Freyhult, A. Rej, M. Staudacher,
{\sl A Generalized Scaling Function for AdS/CFT},
J. Stat. Mech. {\bf 07} (2008) P015 and
arXiv:0712.2743 [hep-th];
\bibitem{FRS2}
A. Belitsky, G. Korchemsky, R. Pasechnik,
{\sl Fine structure of anomalous dimensions in N=4 super Yang-Mills theory},
Nucl. Phys. {\bf B809} (2009) 244 and
arXiv:0806.3657 [hep-ph];
\bibitem{FRS3}
D. Fioravanti, P. Grinza, M. Rossi, {\sl Strong coupling for planar
 ${\cal N}=4$ SYM: an all-order result},  Nucl. Phys. {\bf B810} (2009) 563
and arXiv:0804.2893 [hep-th];
\bibitem{FRS4}
B. Basso, G.P. Korchemsky, {\sl Embedding nonlinear $O(6)$ sigma
model into ${\cal N}=4$ super-Yang-Mills theory},
Nucl. Phys. {\bf B807}(2009) 397 and arXiv:0805.4194 [hep-th];
\bibitem{FRS5}
D. Fioravanti, P. Grinza, M. Rossi,
{\sl The generalised scaling function: a note},
Nucl. Phys. {\bf B827} (2010) 359 and arXiv:0805.4407 [hep-th];
\bibitem{FRS6}
D. Fioravanti, P. Grinza and M. Rossi,
{\sl The generalised scaling function: a systematic study},
JHEP{\bf 11} 2009 037 and arXiv:0808.1886 [hep-th];

\bibitem{FMQR}
D. Fioravanti, A. Mariottini, E. Quattrini, F. Ravanini, {\sl
Excited state Destri-de Vega equation for sine-Gordon and restricted
sine-Gordon models}, Phys. Lett. {\bf B390} (1997) 243
and hep-th/9608091;


\bibitem{FRXYZ}
D. Fioravanti, M. Rossi, {\sl From finite geometry exact
quantities to (elliptic) scattering amplitudes for spin chains:
the 1/2-XYZ}, JHEP{\bf 08}(2005), 010 and hep-th/0504122;


\bibitem {ES}
B. Eden, M. Staudacher,
{\sl Integrability and transcendentality},
J. Stat. Mech. {\bf 11} (2006) P014 and hep-th/0603157;

\bibitem{FTT}
S. Frolov, A. Tirziu, A. Tseytlin,
{\sl Logarithmic corrections to higher twist scaling at strong coupling from AdS/CFT},
Nucl. Phys. {\bf B766} (2007) 232 and hep-th/0611269;

\bibitem{BBKS1}
M. K. Benna, S. Benvenuti, I. R. Klebanov, A. Scardicchio, {\sl A
Test of the AdS/CFT Correspondence Using High-Spin Operators},
Phys. Rev. Lett. {\bf 98} (2007) 131603 and hep-th/0611135;
\bibitem{BBKS2}
L. F. Alday, G. Arutyunov, M. K. Benna, B. Eden, I. R. Klebanov,
{\sl On the Strong Coupling Scaling Dimension of High Spin
Operators}, JHEP{\bf 04} (2007) 082 and hep-th/0702028;
\bibitem{BBKS3}
I. Kostov, D. Serban and D. Volin, {\sl Strong coupling
limit of Bethe Ansatz equations},  Nucl.
Phys. {\bf B789} (2008) 413 and hep-th/0703031;
\bibitem{BBKS4}
M. Beccaria, G.F. De Angelis, V. Forini, {\sl The scaling function
at strong coupling from the quantum string Bethe equations},
JHEP{\bf 04} (2007) 066 and hep-th/0703131;

\bibitem {CK1}
P.Y. Casteill, C. Kristjansen,
{\sl The strong coupling limit of
the scaling function from the quantum string Bethe Ansatz},
Nucl. Phys. {\bf B785} (2007) 1 and arXiv:0705.0890 [hep-th];
\bibitem {CK2}
N. Gromov,
{\sl Generalized Scaling Function at Strong Coupling},
JHEP{\bf 11} 085 and arXiv:0805.4615 [hep-th];

\bibitem{BKK1}
B. Basso, G. Korchemsky, J. Kotanski,
{\sl Cusp anomalous dimension in maximally supersymmetric Yang-Mills theory at strong
coupling},
Phys. Rev. Lett. {\bf 100} (2008) 091601 and arXiv:0708.3933 [hep-th];
\bibitem{BKK2}
I. Kostov, D. Serban and D. Volin,
{\sl Functional BES equation},
JHEP{\bf 08} (2008) 101 and arXiv:0801.2542 [hep-th];

\bibitem{FZ}
L. Freyhult, S. Zieme,
{\sl The virtual scaling function of AdS/CFT },
Phys. Rev. {\bf D 79} (2009) 105009 and
arXiv:0901.2749 [hep-th];

\bibitem{FGR4}
D. Fioravanti, P. Grinza, M. Rossi,
{\sl Beyond cusp anomalous dimension from integrability},
Phys. Lett. {\bf B675} (2009) 137
and arXiv:0901.3161 [hep-th];

\bibitem{FIR}
D. Fioravanti, G. Infusino and M. Rossi,
{\sl On the high spin expansion in the $sl(2)$ ${\cal N}=4$ SYM theory},
Nucl. Phys. {\bf B822} (2009) 467 and
arXiv:0901.3147 [hep-th];

\bibitem{BDM1}
V.M. Braun, S.E. Derkachov, A.N. Manashov, {\sl Integrability of
three particle evolution equations in QCD}, Phys. Rev. Lett. {\bf
81} (1998) 2020 and hep-ph/9805225;
\bibitem{BDM2}
V.M. Braun, S.E. Derkachov, G.P. Korchemsky, A.N. Manashov, {\sl Baryon distribution amplitudes in QCD}, Nucl. Phys. {\bf B553} (1999) 355 and hep-ph/9902375;
\bibitem{BDM3}
A.V. Belitsky, {\sl Fine structure of spectrum of twist-three operators in QCD}, Phys. Lett. {\bf B453} (1999) 59 and hep-ph/9902361;
\bibitem{BDM4}
A.V. Belitsky,  {\sl Integrability and WKB solution of twist-three evolution equations},
Nucl.Phys. {\bf B558} (1999) 259 and hep-ph/9903512;
\bibitem{BDM5}
A.V. Belitsky, {\sl Renormalization of twist-three operators and
integrable lattice models},  Nucl. Phys. {\bf B574} (2000) 407 and hep-ph/9907420;
\bibitem{BDM6}
A.V. Belitsky, A.S. Gorsky, G.P. Korchemsky, {\sl Gauge/string
duality for QCD conformal operators}, Nucl. Phys. {\bf B667} (2003)
3 and hep-th/0304028;

\bibitem{BKM1}
A.V. Belitsky, G.P. Korchemsky, D. Mueller,
{\sl Towards Baxter equation in supersymmetric Yang-Mills theories},
Nucl. Phys. {\bf B768} (2007) 116 and  hep-th/0605291;
\bibitem{BKM2}
A.V. Belitsky,
{\sl Long-range SL(2) Baxter equation in N=4 super-Yang-Mills theory},
Phys. Lett. {\bf B643} (2006) 354 and
hep-th/0609068;

\bibitem{Mat1}
D.V. Boulatov,
{\sl Wilson loop on a sphere},
Mod.Phys.Lett. A9 (1994) 365 and hep-th/9310041;
\bibitem{Mat2}
J.M. Daul, V. Kazakov,
{\sl Wilson loop for large N Yang-Mills theory on a two-dimensional sphere},
Phys.Lett. {\bf B335} (1994) 371 and hep-th/9310165;

\bibitem{BTZ1}
N. Beisert, A. Tseytlin, K. Zarembo,
{\sl Matching quantum strings to quantum spins: One-loop versus finite-size corrections},
Nucl.Phys. {\bf B715} (2005) 190 and hep-th/0502173;
\bibitem{BTZ2}
R. Hernandez, E. Lopez, A. Perianez, G. Sierra,
{\sl Finite size effects in ferromagnetic spin chains and quantum corrections to classical strings},
JHEP {\bf 06} (2005) 011 and hep-th/0502188;

\bibitem{KOR}
G.P. Korchemsky, {\sl Quasiclassical QCD pomeron}, Nucl. Phys. {\bf B462} (1996) 333 and hep-th/9508025;

\bibitem{FR}
D. Fioravanti, M. Rossi,
{\sl TBA-like equations and Casimir effect in (non-)perturbative AdS/CFT},
JHEP12 (2012) 013 and arXiv:1112.5668 [hep-th];

\bibitem{rep}
D. Fioravanti, M. Rossi,
{\sl The high spin expansion of twist sector dimensions: the planar ${\cal N}=4$ super Yang-Mills theory},
Adv. High Energy Phys.  {\bf 2010} (2010)  614130 and arXiv:1004.1081 [hep-th];

\bibitem{KM1}
G.P. Korchemsky,
{\sl Asymptotics of the Altarelli-Parisi-Lipatov Evolution Kernels of Parton Distributions},
Mod.Phys.Lett. {\bf A4} (1989) 1257;
\bibitem{KM2} 
G.P. Korchemsky, G. Marchesini,
{\sl Structure function for large x and renormalization of Wilson loop},
Nucl.Phys. {\bf B406} (1993) 225 and hep-ph/9210281;

\bibitem{BFTT}
M. Beccaria, V. Forini, A. Tirziu, A.A. Tseytlin,
{\sl Structure of large spin expansion of anomalous dimensions at strong coupling},
Nucl. Phys. {\bf B812} (2009) 144 and
arXiv:0809.5234 [hep-th];

\bibitem{KL}
A.V. Kotikov, L.N. Lipatov,
{\sl On the highest transcendentality in N=4 SUSY},
Nucl. Phys. {\bf B769} (2007) 217 and hep-th/0611204;

\bibitem{FGR5}
D. Fioravanti, P. Grinza, M. Rossi, {\sl On the logarithmic powers of $sl(2)$ SYM$_4$},
Phys. Lett. {\bf B684} (2010) 52 and arXiv:0911.2425 [hep-th];

\bibitem{BM}
G. Macorini, M. Beccaria, {\sl Reciprocity of higher conserved charges in the ${sl}(2)$ sector of
${\cal N}=4$ SYM}, arXiv:1009.5559[hep-th];

\bibitem{GSSV1}
M. Beccaria, G.V. Dunne, V. Forini, M. Pawellek, A.A. Tseytlin,
{\sl Exact computation of one-loop correction to energy of spinning folded string in $AdS_5 x S^5$},
J.Phys. {\bf  A43} (2010) 165402 and arXiv:1001.4018 [hep-th];
\bibitem{GSSV2}
S. Giombi, R. Ricci, R. Roiban, A.A. Tseytlin,
{\sl Two-loop $AdS_5 x S^5$ superstring: testing asymptotic Bethe ansatz and finite size corrections},
J.Phys. {\bf  A44} (2011) 045402 and arXiv:1010.4594 [hep-th];
\bibitem{GSSV3}
N. Gromov, D. Serban, I. Shenderovich, D. Volin,
{\sl Quantum folded string and integrability: From finite size effects to Konishi dimension},
JHEP {\bf 08} (2011) 046 and arXiv: 1102.1040 [hep-th];
\bibitem{GSSV4}
B. Basso, A.A. Belitsky,
{\sl L\"{u}scher formula for the GKP string},
Nucl. Phys. {\bf B860} (2012) 1
and arXiv:1108.0999 [hep-th];

\bibitem{BC}
M. Beccaria,
{\sl Anomalous dimensions at twist-3 in the sl(2) sector of N=4 SYM},
JHEP {\bf 06} (2007) 044 and arXiv:0704.3570 [hep-th].














\end{thebibliography}
\end{document}